\documentclass{article}

\usepackage{survey_style}
\usepackage[utf8]{inputenc}
\usepackage[T1]{fontenc}

\usepackage{newunicodechar}
\newunicodechar{±}{\ensuremath{\pm}}
\newunicodechar{×}{\ensuremath{\times}}
\newunicodechar{→}{\ensuremath{\rightarrow}}
\newunicodechar{≤}{\ensuremath{\leq}}
\newunicodechar{≥}{\ensuremath{\geq}}

\usepackage[numbers,sort]{natbib}

\usepackage{graphicx}
\usepackage{booktabs}
\usepackage{amsfonts}
\usepackage{amssymb}
\usepackage{amsmath}
\usepackage{amsthm}
\usepackage{microtype}
\usepackage{nicefrac}
\usepackage{xspace}
\usepackage{float}
\usepackage{url}
\usepackage{multirow}
\usepackage{subcaption}
\usepackage{makecell}
\usepackage{tabularx}
\usepackage{array}
\usepackage{textcomp}
\usepackage[table,xcdraw,dvipsnames]{xcolor}
\usepackage[most]{tcolorbox}
\usepackage{enumitem}
\usepackage{caption}
\DeclareCaptionLabelSeparator{pipe}{ | }
\usepackage{algorithm}
\usepackage{algpseudocode}

\usepackage{hyperref}
\usepackage{cleveref}
\hypersetup{
  colorlinks=true,
  linkcolor=blue,
  urlcolor=blue,
  citecolor=blue,
  linkbordercolor=blue,
  urlbordercolor=blue,
  citebordercolor=blue,
  pdfborderstyle={/S/U/W 1},
}

\providecommand{\IEEEPARstart}[2]{#1#2}

\theoremstyle{plain}

\theoremstyle{definition}

\theoremstyle{remark}

\definecolor{abstractbg}{HTML}{F0F0F0}
\newtcolorbox{abstractbox}{
  enhanced,
  breakable,
  colback=abstractbg,
  colframe=abstractbg,
  boxrule=0pt,
  arc=2.5mm,
  left=3.5mm,
  right=3.5mm,
  top=2.5mm,
  bottom=2.5mm,
  before skip=8pt,
  after skip=12pt
}
\renewenvironment{abstract}
{
  \begin{abstractbox}
  \begin{center}
    {\large\bfseries Abstract}
  \end{center}
  \vspace{-0.25\baselineskip}
}
{
  \par\medskip
  \end{abstractbox}
}

\setlist[itemize]{leftmargin=*}
\setlist[enumerate]{leftmargin=*}
\setlist[description]{leftmargin=*}

\makeatletter
\renewcommand{\@institutionlogos}{%
  \noindent
  \begin{minipage}[c][0.46in][c]{0.465\linewidth}
    \raggedright
    \includegraphics[height=0.32in,keepaspectratio]{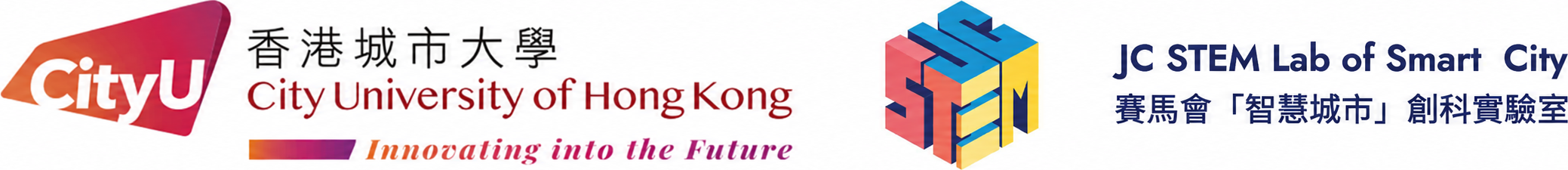}%
  \end{minipage}%
  \hfill
  \begin{minipage}[c][0.46in][c]{0.525\linewidth}
    \raggedleft
    \includegraphics[height=0.32in,keepaspectratio]{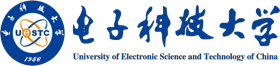}%
    \hspace{0.12in}%
    \includegraphics[height=0.32in,keepaspectratio]{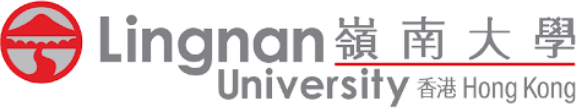}%
  \end{minipage}\par
  \vskip -0.10in
}
\makeatother

\title{CP-Guard+: A New Paradigm for Malicious Agent Detection and Defense in Multi-Agent Collaborative Perception}

\renewcommand{\shorttitle}{CP-Guard+}

\author{
\bfseries Senkang Hu\textsuperscript{1,2}, Yihang Tao\textsuperscript{1,2}, Zihan Fang\textsuperscript{1,2}, Guowen Xu\textsuperscript{3}, Hangcheng Cao\textsuperscript{1,2}, \\
\bfseries Yiqin Deng\textsuperscript{4}, Sam Tak Wu Kwong\textsuperscript{4}, Yuguang Fang\textsuperscript{1,2} \\
\textsuperscript{1}Hong Kong JC STEM Lab of Smart City,
\textsuperscript{2}City University of Hong Kong, \\
\textsuperscript{3}University of Electronic Science and Technology of China,
\textsuperscript{4}Lingnan University. \\
{\small\texttt{\{senkang.forest, yihang.tommy, zihanfang3-c\}@my.cityu.edu.hk}, \texttt{\{hangccao, my.Fang\}@cityu.edu.hk},} \\
{\small\texttt{guowen.xu@uestc.edu.cn}, \texttt{\{yiqindeng, samkwong\}@ln.edu.hk}} \\
}

\begin{document}
\maketitle

{\renewcommand{\thefootnote}{}%
\footnotetext{The research work described in this paper was conducted in the JC STEM Lab of Smart City funded by The Hong Kong Jockey Club Charities Trust under Contract 2023-0108. The work was supported in part by the Hong Kong SAR Government under the Global STEM Professorship and Research Talent Hub. The work of S. Hu was supported in part by the Hong Kong Innovation and Technology Commission under InnoHK Project CIMDA. The work of Y. Deng was supported in part by the National Natural Science Foundation of China under Grant No. 62301300. The work of G. Xu was supported by the National Natural Science Foundation of China under Grant 62502075.}}

\begin{abstract}
    Collaborative perception (CP) is a promising approach for safe connected and autonomous driving that enables multiple connected and autonomous vehicles (CAVs) to share sensing information with each other to enhance perception performance. For example, occluded objects can be detected, and the sensing range can be extended. However, compared with single-agent perception, the openness of a CP system makes it more vulnerable to malicious agents and attackers, who can inject malicious information to mislead the perception of an ego CAV, resulting in severe risks for the safety of autonomous driving systems. To mitigate the vulnerability of CP systems, we first propose a new paradigm for malicious agent detection that effectively identifies malicious agents at the feature level without requiring verification of final perception results, significantly reducing computational overhead. Building on this paradigm, we introduce CP-GuardBench, the first comprehensive dataset for training and evaluating various malicious agent detection methods for CP systems. Furthermore, we develop a robust defense method called CP-Guard+, which enhances the margin between the representations of benign and malicious features through a carefully designed dual-centered contrastive loss. Finally, we conduct extensive experiments on both CP-GuardBench and V2X-Sim, and the results demonstrate the superiority of CP-Guard+.
\end{abstract}

\keywords{Connected and Autonomous Vehicles, Collaborative Perception, Malicious Agent Defense, Multi-Agent System.}

\section{Introduction}
\label{sec:intro}

\IEEEPARstart{T}{he} development of collaborative perception (CP) has been driven by the increasing demand for accurate and reliable perception in autonomous driving systems \citep{chenCooperCooperativePerception2019,chenFcooperFeatureBased2019,liV2XSimMultiAgentCollaborative2022,huWhere2commCommunicationefficientCollaborative2024,huAdaptiveCommunicationsCollaborative2023,huFullSceneDomainGeneralization2024,fangPACPPriorityAwareCollaborative2024,xuOPV2VOpenBenchmark2022}. Single-agent perception systems, which rely solely on the onboard sensors of a single vehicle, are restricted by a limited sensing range and occlusion. In contrast, CP systems incorporate multiple connected and autonomous vehicles (CAVs) to collaboratively capture their surrounding environments. Specifically, CAVs in a CP system can be divided into two categories: the ego CAV and helping CAVs. The helping CAVs send complementary sensing information (most methods send intermediate features) to the ego CAV, which then leverages this complementary information to enhance its perception performance \citep{balkusSurveyCollaborativeMachine2022,hanCollaborativePerceptionAutonomous2023,huAgentsCoDriverLargeLanguage2024,wangV2VNetVehicletoVehicleCommunication2020a}. For example, the ego CAV can detect occluded objects and extend its sensing range after fusing the received information.

\begin{figure*}[htbp]
    \centering

    \begin{subfigure}[b]{0.504\linewidth}
        \includegraphics[width=\textwidth]{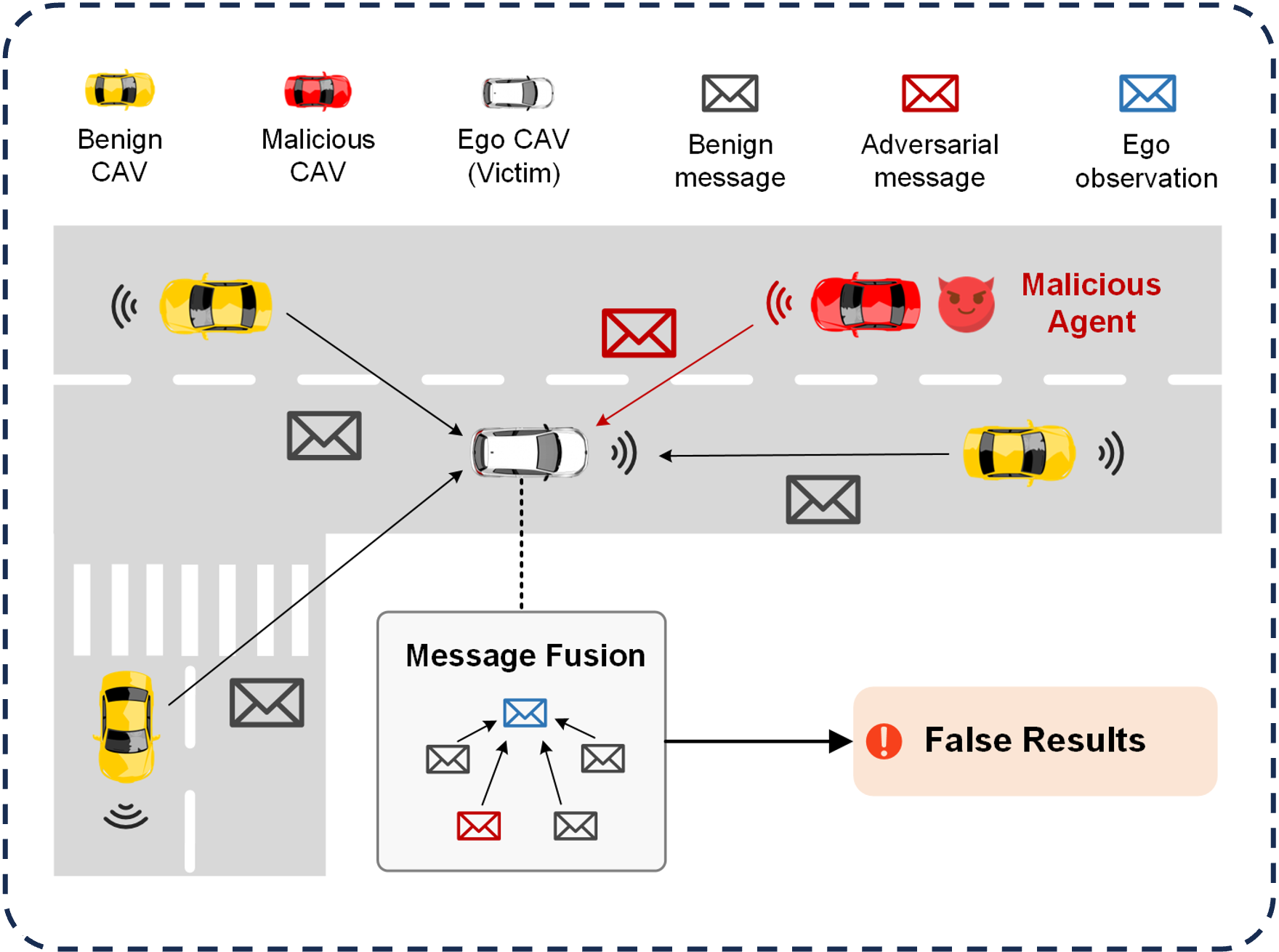}
        \caption{Threats in Collaborative Perception}
        \label{fig:security_threats}
    \end{subfigure}
    \hfill
    \begin{subfigure}[b]{0.45\linewidth}
        \includegraphics[width=\textwidth]{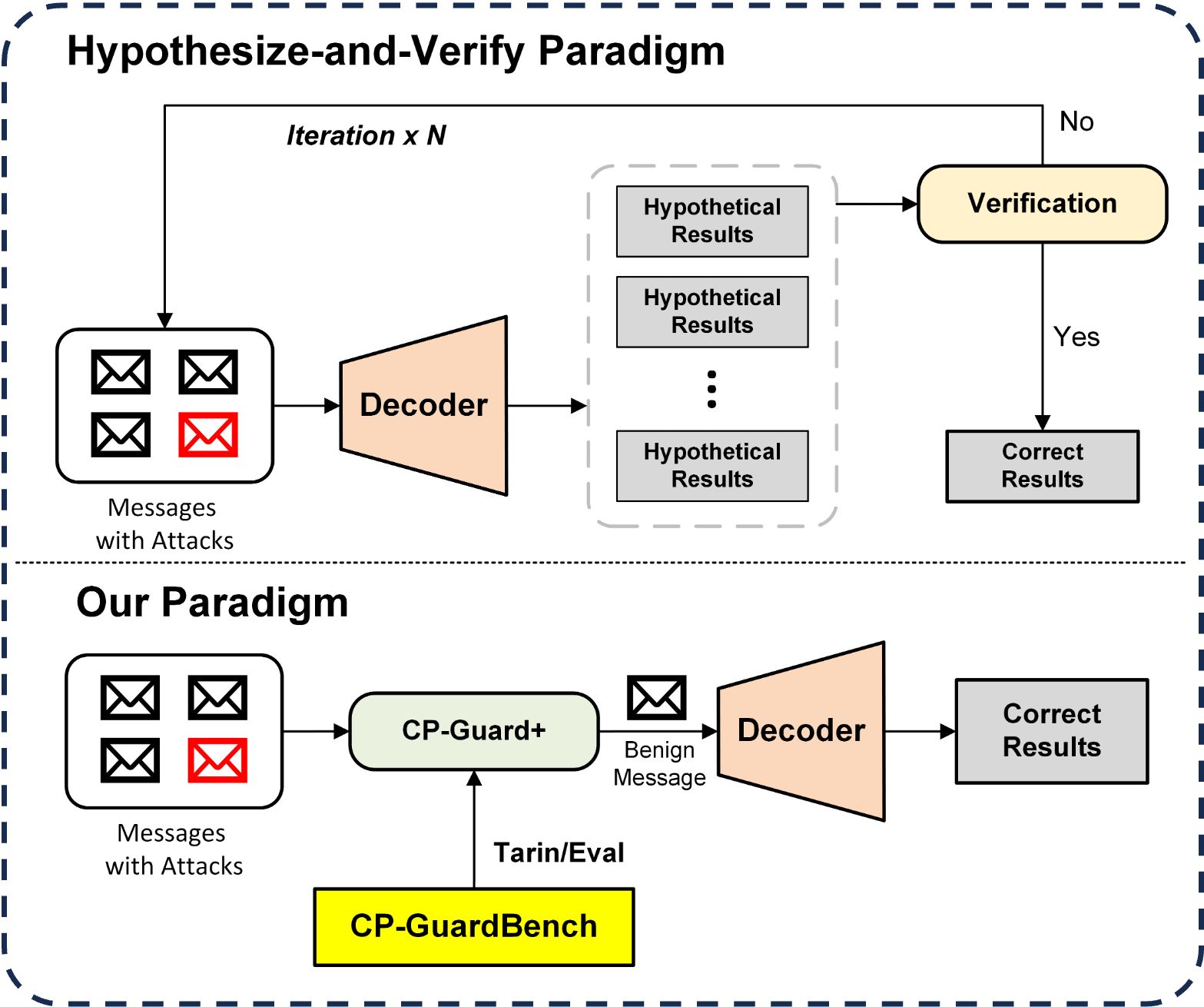}
        \caption{Paradigm Comparison}
        \label{fig:idea_comparison}
    \end{subfigure}
    \caption{(a) \textbf{Illustration of the threats of malicious agents in collaborative perception}. Malicious CAVs could send intricately crafted adversarial messages to an ego CAV, which can mislead it into generating false positive perception outputs. (b) \textbf{Comparison between the proposed CP-Guard+ with the traditional hypothesize-and-verify malicious agent detection methods.}
    Hypothesize-and-verify involves multiple rounds of malicious agent detection iterations at the output level and requires the generation of multiple hypothetical outputs for verification, incurring high computational overhead. In contrast, CP-Guard+ directly outputs robust CP results with intermediate feature-level detection, significantly reducing the computational overhead.}
    \vspace{-5mm}
\end{figure*}

Despite many advantages of CP outlined above, it also has several significant limitations. Compared to single-agent perception systems, CP is more vulnerable to security threats and easier to attack since it requires receiving and fusing information from other CAVs, which expands the attack surface. In particular, malicious agents can directly send intermediate features with adversarial perturbations to fool the ego CAV, or a man-in-the-middle attacker can capture and manipulate the intermediate feature maps. Figure \ref{fig:security_threats} illustrates the vulnerability of CP to malicious agents. Additionally, several attack methods have been designed to fool CP. For example, Tu \textit{et al.} \citep{tuAdversarialAttacksMultiAgent2021} developed a method to generate indistinguishable adversarial perturbations to attack multi-agent communication in CP, which can severely degrade perception performance.

The ego CAV's inability to accurately detect and eliminate malicious agents from its collaboration network poses significant risks to CP, potentially resulting in compromised perception outcomes with catastrophic consequences.
For instance, the ego CAV might misinterpret traffic light statuses or fail to detect objects ahead on the road, leading to severe traffic accidents or even fatalities. Hence, it is crucial to develop a defense mechanism for CP that is resilient to attacks from malicious agents and capable of eliminating them from its collaboration network.

To address the security threats in CP, some prior studies have investigated defense mechanisms against malicious agents. For example, Li \textit{et al.} \citep{liUsAdversariallyRobust2023} leveraged random sample consensus (RANSAC) to sample a subset of collaborators and calculate the intersection over union (IoU) of the bounding boxes to verify whether there are any malicious agents within the collaboration network. Zhao \textit{et al.} \citep{zhaoMaliciousAgentDetection2024} designed a match loss and a reconstruction loss as statistics to measure the consensus between the ego CAV and its collaborators.
However, these methods all follow a hypothesize-and-verify paradigm, which requires generating multiple hypothetical perception results and verifying the consistency between the ego CAV and the collaborators.
This approach is computationally intensive and time-consuming, which hinders its scalability and leads to the following question:
\begin{quote}
    \centering
    \textit{Is it feasible to detect malicious agents directly at the feature level?}
\end{quote}
As illustrated in Figure \ref{fig:idea_comparison}, the new paradigm shifts the focus to feature-level detection, eliminating the need to generate multiple hypothetical perception results. This direct approach can significantly reduce computational overhead for malicious agent detection in CP systems.

Although this idea is concise and appealing, there are still some challenges in realizing it. Firstly, to detect malicious agents at the feature level, we need to train a deep neural network (DNN) model on a large-scale dataset to help it learn the features of benign and malicious agents. However, no benchmark dataset exists for feature-level malicious agent detection in CP systems. The existing datasets for CP, such as V2X-Sim \citep{liV2XSimMultiAgentCollaborative2022} and OPV2V \citep{xuOPV2VOpenBenchmark2022}, contain only benign agents and do not include malicious agents.
Therefore, it is difficult to train a robust DNN model for malicious agent detection in CP systems without a well-annotated dataset.
Secondly, in CP scenarios, the environments are highly dynamic and complex, making it infeasible to directly use a classifier to categorize the received intermediate features for detecting malicious agents. The reason is that dynamic environments can lead to a high false-positive rate (FPR). Additionally, adversarial perturbations are indistinguishable at the feature level, and the feature distributions of malicious agents and benign agents are highly similar. These factors make it difficult to train a robust model to distinguish malicious agents from benign agents.

To address the aforementioned challenges, we first generate a new dataset, CP-GuardBench, which is the first dataset for malicious agent detection in CP systems. Then, we propose CP-Guard+, a robust malicious agent detection method for CP systems.
CP-Guard+ can effectively detect malicious agents at the feature level without verifying the final perception results, significantly reducing computational overhead and improving defense efficiency. Moreover, we design a Dual-Centered Contrastive Loss (DCCLoss) to tackle indistinguishability issues at the feature level and further enhance robustness.

In summary, we investigate the malicious agent detection problem in CP systems and propose a new paradigm, namely, feature-level malicious agent detection. Additionally, we construct CP-GuardBench, the first dataset for malicious agent detection in CP systems. Furthermore, we propose CP-Guard+, a robust malicious agent detection framework with high robustness and computational efficiency. Finally, we conduct extensive experiments on CP-GuardBench and V2X-Sim and demonstrate the superiority of our CP-Guard+.

\section{Related Work}
\label{sec:related_work}

\subsection{Collaborative Perception}
Collaborative Perception (CP) significantly extends the field-of-view (FoV) for individual agents, thereby enhancing the comprehensiveness and accuracy of perception outcomes \citep{hanCollaborativePerceptionAutonomous2023,huCollaborativePerceptionConnected2024}. In CP systems, CAVs utilize various fusion methods tailored at distinct stages of data processing. Early fusion at the raw data level and late fusion at the output level often result in either high communication loads or increased perceptual noise. Conversely, intermediate fusion, which involves the transmission of intermediate features among CAVs, achieves an optimal balance by minimizing communication overhead while maximizing perceptual accuracy.
Based on intermediate-level collaboration, recent progress in CP has addressed a wide array of challenges, including communication overhead \citep{fangPACPPriorityAwareCollaborative2024, tao2024directedcpdirectedcollaborativeperception}, robustness \citep{10160546}, system heterogeneity \citep{lu2024an}, and domain generalization \citep{huFullSceneDomainGeneralization2024}. Robustness, in particular, has become a pivotal area of focus, tackling issues such as communication disruptions \citep{10457955}, pose noise correction \citep{10160546}, and system latency \citep{10.1007/978-3-031-19824-3_19}. Despite comprehensive research, the vulnerability of these systems to malicious attacks has not been adequately addressed. In this paper, we investigate the robustness of CP systems, specifically considering the impact of malicious agents, and propose strategies to enhance system security and integrity.

\subsection{Adversarial Collaborative Perception}

Adversarial attacks on single-vehicle perception systems typically utilize methods such as GPS spoofing \citep{Li_2021_ICCV}, LiDAR spoofing \citep{279980}, and deploying physically realizable adversarial objects \citep{Tu_2020_CVPR}. However, in multi-vehicle collaborative perception, adversarial strategies differ markedly across collaboration stages.
For early-stage collaborative perception, Zhang \textit{et al.} \citep{zhangDataFabricationCollaborative2023} have identified attacks that involve object spoofing and removal, exploiting vulnerabilities through simulated object presence or absence and advanced reconstruction of LiDAR point clouds. Conversely, late-stage collaboration, which mainly involves sharing object locations \citep{9120490}, offers adversaries opportunities to manipulate these data points.
Attacks at the intermediate stage are more complex, typically requiring white-box access to perception models. Such access allows attackers to precisely manipulate system outputs, though these systems are generally less vulnerable to simple black-box strategies like ray-casting due to the protective nature of benign feature maps that diminish the impact of these attacks.
Pioneering work by Tu \textit{et al.} \citep{tuAdversarialAttacksMultiAgent2021} introduced untargeted adversarial attacks aimed at generating inaccurate detection bounding boxes by altering feature maps in intermediate-fusion systems. Building on this, Zhang \textit{et al.}~\citep{zhangDataFabricationCollaborative2023} have enhanced these techniques by incorporating perturbation initialization and feature map masking, enabling more realistic and targeted attacks in real-time scenarios.
Our research focuses on identifying and mitigating adversarial threats within the intermediate-level collaborative perception framework to bolster system resilience against these sophisticated attacks.

\subsection{Defensive Collaborative Perception}

To enhance the resilience of intermediate-level CP against adversarial threats, contemporary research has primarily focused on the detection of malicious agents at the output level. Li \textit{et al.} \citep{liUsAdversariallyRobust2023} developed the Robust Collaborative Sampling Consensus (ROBOSAC) method that selects a random subset of collaborators for consensus verification. Additionally, Zhao \textit{et al.} \citep{zhaoMaliciousAgentDetection2024} introduced match loss and reconstruction loss as metrics to assess consensus between an ego CAV and its collaborators' perception results for the detection of malicious agents. Furthermore, Zhang \textit{et al.} \citep{zhangDataFabricationCollaborative2023} utilized occupancy maps to identify inconsistencies between an ego CAV and other collaborators. In addition, CP-Guard \cite{hu2024cpguardmaliciousagentdetection} leverages the collaborative bird's eye view (BEV) segmentation results to iteratively check the normality from different collaborators to defend against malicious agents.
However, these approaches adhere to a hypothesize-and-verify workflow, necessitating the generation of hypothetical perception outcomes and subsequent verification of their consistency with those of collaborators. This methodology is notably time-consuming and resource-intensive, hindering the system scalability. In this paper, we have proposed a novel approach that shifts the focus to detecting malicious agents at the feature level, thus circumventing the need to verify final perception results. In addition, Zhang \textit{et al.} \cite{zhangCooperativePerceptionSafe2023} leverage collaborative perception to defend LiDAR spoofing attack. Specifically, the authors use LiDAR scan data from neighboring vehicles to help the ego vehicle to detect and mitigate LiDAR spoofing attacks. Since current spoofing hardware typically targets one vehicle at a time, comparison with other vehicles' data helps identify discrepancies. This method essentially defends against conventional threats (LiDAR spoofing attack, GPS attack, etc.) to autonomous driving, while using collaborative perception as a means of detecting attacks. In contrast, our method focuses on the threats specific to the collaborative perception system itself, rather than leveraging collaborative perception as a means to address general vehicular threats.

\subsection{Anomaly Detection}

Malicious agent detection in CP systems can be viewed as a specific instance of anomaly detection, as it involves distinguishing abnormal agents from normal ones. This section provides a concise overview of related work in anomaly detection. Recent advancements in this field have utilized self-supervised and contrastive learning to enhance representation learning for differentiating between normal and anomalous data. For example, Masked Contrastive Learning (MCL) \cite{choMaskedContrastiveLearning2021} improves SimCLR by incorporating a Class-Conditional Mask (CCM) to adjust the repelling ratio based on label information and Stochastic Positive Attraction (SPA) to enhance cluster compactness, creating dense class-conditional clusters without the need for fine-tuning. Similarly, the Mean-Shifted Contrastive Loss (MSC) \cite{reissMeanShiftedContrastiveLoss2023} modifies pre-trained features for anomaly detection by shifting representations towards the center of normal data, ensuring compact clusters and improved separation of anomalies, achieving state-of-the-art results on benchmarks like CIFAR-10 and CIFAR-100. In contrast, CADet \cite{guille-escuretCADetFullySelfsupervised2024} offers a fully self-supervised approach by integrating contrastive learning with MMD-inspired statistics to detect unseen classes and adversarial samples without requiring labels or out-of-distribution (OOD) examples. In the context of vehicle-related applications, Kopuklu \textit{et al.} \cite{kopukluDriverAnomalyDetection2021a} introduced a dataset and contrastive learning method to identify anomalies in driver behaviors (e.g., talking, drinking) that could lead to accidents. Jiao \textit{et al.} \cite{jiaoLearningRepresentationAnomaly2023} employ contrastive learning and an unsupervised method using reconstruction with semantic latent space to detect anomalies in vehicle trajectories. Additionally, Han \textit{et al.} \cite{hanADSLeadLifelongAnomaly2023} developed ADS-Lead to detect GPS spoofing attacks (time series anomalies) and traffic sign and lane recognition attacks. These studies highlight significant advancements in anomaly detection and its applications.

\begin{algorithm}[t]
    \caption{Attack Process on Collaborative Perception}
    \label{alg:attack_process}
    \begin{algorithmic}[1]
    \Require Feature encoder $f_\mathtt{encoder}$, feature aggregator $f_\mathtt{agg}$, feature decoder $f_\mathtt{decoder}$; raw observations $\{\mathbf{O}_k\}_{k \in \mathcal{X}^N}$ ($\mathbf{O}_k \in \mathbb{R}^{H_0 \times W_0 \times 4}$, raw LiDAR BEV voxels); perturbation bound $\Delta$; learning rate $\alpha$; iterations $T$
    \Ensure Adversarial perturbation $\delta^* \in \mathbb{R}^{H \times W \times C}$ (same shape as $\mathbf{F}_k$; in our implementation $H\!=\!W\!=\!32$, $C\!=\!256$, single-agent processing: batch size $B\!=\!1$ per CAV)

    \State // {Phase 1: Local Perception}
    \For{each CAV $k \in \mathcal{X}^N$ \textbf{in parallel}}
        \State $\mathbf{F}_k \gets f_\mathtt{encoder}(\mathbf{O}_k)$ \Comment{Extract intermediate features; $\mathbf{F}_k \in \mathbb{R}^{H \times W \times C}$}
    \EndFor

    \State // {Phase 2: Feature Communication}
    \For{each CAV $j \in \mathcal{X}^N, j \neq i$}
        \State Transmit $\mathbf{F}_{j \rightarrow i} \gets \mathbf{\Gamma}_{j \rightarrow i}(\mathbf{F}_j)$ to ego CAV $i$ \Comment{$\mathbf{F}_{j\rightarrow i} \in \mathbb{R}^{H \times W \times C}$ after spatial alignment}
    \EndFor

    \State // {Phase 3: Attack Generation}
    \State For malicious CAV $m$, initialize $\delta^0 \gets \mathbf{0}$
    \For{$t = 0$ to $T-1$}
        \State $\mathbf{F}_m^\delta \gets \mathbf{F}_m + \delta^t$ \Comment{Apply perturbation}
        \State $\mathbf{F}_{m \rightarrow i}^\delta \gets \mathbf{\Gamma}_{m \rightarrow i}(\mathbf{F}_m^\delta)$ \Comment{Coordinate transformation}
        \State $\mathbf{F}_\mathtt{fused}^\delta \gets f_\mathtt{agg}(\mathbf{F}_{i}, \{\mathbf{F}_{j \rightarrow i}\}_{j \neq i, j \neq m}, \mathbf{F}_{m \rightarrow i}^\delta)$ \Comment{Fusion}
        \State $\mathbf{Y}^\delta \gets f_\mathtt{decoder}(\mathbf{F}_\mathtt{fused}^\delta)$ \Comment{Generate perturbed output; $\mathbf{Y}^\delta$: the set of predicted bounding boxes}
        \State Compute $\mathcal{L}(\mathbf{Y}^\delta, \mathbf{Y}^\mathtt{gt})$ \Comment{Calculate adversarial loss}
        \State $\delta^{t+1} \gets \delta^t + \alpha \cdot \nabla_\delta \mathcal{L}(\mathbf{Y}^\delta, \mathbf{Y}^\mathtt{gt})$ \Comment{Update perturbation}
        \State $\delta^{t+1} \gets \text{Clip}(\delta^{t+1}, -\Delta, \Delta)$ \Comment{Enforce perturbation bound}
    \EndFor
    \State $\delta^* \gets \delta^T$ \Comment{Final adversarial perturbation}

    \State // {Phase 4: Final Perception}
    \State Malicious CAV transmits $\mathbf{F}_{m \rightarrow i}^{\delta^*} \gets \mathbf{\Gamma}_{m \rightarrow i}(\mathbf{F}_m + \delta^*)$ to ego CAV
    \State Ego CAV performs feature aggregation: $\mathbf{F}_\mathtt{fused} \gets f_\mathtt{agg}(\mathbf{F}_{i}, \{\mathbf{F}_{j \rightarrow i}\}_{j \neq i, j \neq m}, \mathbf{F}_{m \rightarrow i}^{\delta^*})$
    \State Ego CAV generates final perception: $\mathbf{Y} \gets f_\mathtt{decoder}(\mathbf{F}_\mathtt{fused})$
    \end{algorithmic}
\end{algorithm}

\begin{table*}[t]
    \centering
    \caption{\textbf{Summary of Key Notation.}}
    \vspace{-2mm}
    \label{tab:notation}
    \renewcommand\arraystretch{1.1}
    \begin{tabular}{ll}
        \toprule
        \textbf{Symbol} & \textbf{Description} \\
        \midrule
        $\mathcal{X}^N$ & Set of $N$ CAVs in the CP system \\
        $i, j, k$ & CAV indices; $i$ denotes the ego CAV \\
        $\mathbf{O}_k$ & Raw observation of CAV $k$ (LiDAR BEV voxels, $\mathbb{R}^{H_0\times W_0\times 4}$) \\
        $\mathbf{F}_k$ & Intermediate feature map of CAV $k$ ($\mathbb{R}^{H\times W\times C}$; $H\!=\!W\!=\!32$, $C\!=\!256$) \\
        $\mathbf{F}_{j\to i}$ & Feature of CAV $j$ spatially aligned to ego $i$'s coordinate frame \\
        $\mathbf{F}^\mathtt{res}_{j\to i}$ & Residual feature: $\mathbf{F}_i - \mathbf{F}_{j\to i}$ (input to CP-Guard+ classifier) \\
        $\mathbf{F}_\mathtt{fused}$ & Aggregated feature at ego CAV after fusion \\
        $\mathbf{Y}$ & Final perception output (predicted bounding boxes) \\
        $f_\mathtt{encoder}$ & Feature encoder: $\mathbf{O}_k \mapsto \mathbf{F}_k$ \\
        $f_\mathtt{agg}$ & Feature aggregator (fusion module) \\
        $f_\mathtt{decoder}$ & Feature decoder: $\mathbf{F}_\mathtt{fused} \mapsto \mathbf{Y}$ \\
        $\mathbf{\Gamma}_{j\to i}$ & Spatial transformation and transmission operator \\
        $\delta,\,\delta^*$ & Adversarial perturbation / optimized perturbation ($\|\delta\|\leq\Delta$) \\
        $\Delta$ & Perturbation budget \\
        $\alpha$ & DCCLoss weight (or attack step size, by context) \\
        $f_\mathtt{cls}$ & Binary classifier for malicious-agent detection \\
        \bottomrule
    \end{tabular}
    \vspace{-3mm}
\end{table*}

\section{Preliminaries}
\label{sec:preliminaries}
\subsection{Formulation of Collaborative Perception}
\label{sec:formulation}
In this section, we formulate collaborative perception and give the pipeline of our CP system. Specifically, let $\mathcal{X}^N$ denote the set of $N$ CAVs in the CP system. CAVs in $\mathcal{X}$ can be divided into two categories: the ego CAV and helping CAVs. The ego CAV is the one that needs to perceive its surrounding environment, while helping CAVs are the ones that send their complementary sensing information to the ego CAV to help it enhance its perception performance.
Thus, each CAV can serve as either the ego or a helping one, depending on its role in a perception process. We assume that each CAV is equipped with a feature encoder $f_\mathtt{{encoder}}(\cdot)$, a feature aggregator $f_\mathtt{{agg}}(\cdot)$, and a feature decoder $f_\mathtt{{decoder}}(\cdot)$. For the $i$-th CAV in the set $\mathcal{X}$, the raw observation is denoted as $\mathbf{O}_i$ (such as camera images and LiDAR point clouds), and the final perception results are denoted as $\mathbf{Y}_i$. The CP pipeline of the $i$-th CAV can be described as follows.
\begin{enumerate}
    \setlength{\itemsep}{0pt}
    \setlength{\parskip}{0pt}
    \setlength{\parsep}{0pt}
    \item \textit{Observation Encoding}: Each CAV encodes its raw observation $\mathbf{O}_j$ into an initial feature map $\mathbf{F}_j = f_\mathtt{{encoder}}(\mathbf{O}_j)$, where $j \in \mathcal{X}^N$.
    \item \textit{Intermediate Feature Transmission}: Helping CAVs transmit their intermediate features to the ego CAV: $\mathbf{F}_{j\rightarrow i}=\mathbf{\Gamma}_{j\rightarrow i}(\mathbf{F}_j),\  j\in \mathcal{X}^N, j\neq i,$
    where $\mathbf{\Gamma}_{j\rightarrow i}(\cdot)$ denotes a transmitter that conveys the $j$-th CAV's intermediate feature $\mathbf{F}_j$ to the ego CAV, while performing a spatial transformation. $\mathbf{F}_{j\rightarrow i}$ is the spatially aligned feature in the $i$-th CAV's coordinate.
    \item \textit{Feature Aggregation}: The ego CAV receives all the intermediate features and fuses them into a unified observational feature $\mathbf{F}_\mathtt{fused}=f_\mathtt{agg}(\mathbf{F}_{0\rightarrow i}, \{\mathbf{F}_{j\rightarrow i}\}_{j\neq i,\  j\in \mathcal{X}^N})$.
    \item \textit{Perception Decoding}: Finally, the ego CAV decodes the unified observational feature $\mathbf{F}_\mathtt{fused}$ into the final perception results $\mathbf{Y}=f_\mathtt{decoder}(\mathbf{F}_\mathtt{fused})$.
\end{enumerate}

\begin{figure*}[t]
    \centering
    \includegraphics[width=.8\linewidth]{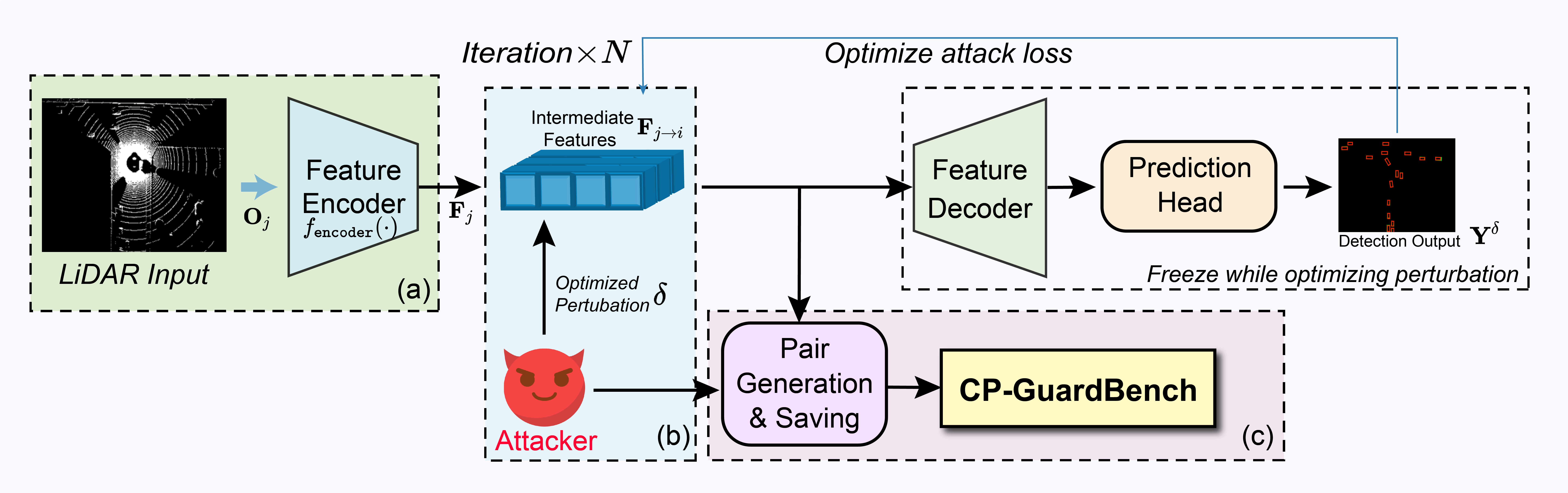}
    \caption{\textbf{Automatic Data Generation and Annotation Pipeline.} We first train a robust LiDAR collaborative object detector. Then, we discard the detection head and decoder and only keep the backbone as the intermediate feature generator. The data generation pipeline is shown in (a), (b), and (c), where (a) is the intermediate feature generation, (b) is the attack implementation, and (c) is the pair generation and saving.}
    \label{fig:data_generation}
    \vspace{-5mm}
\end{figure*}

\subsection{Adversarial Threat Model}

Our focus is on the operation of an intermediate-fusion collaboration scheme, where an attacker introduces designed adversarial perturbations into the intermediate features to mislead the perception of the ego CAV. Since an attacker participates in the collaborative system with local perception model installation, we assume they have white-box access to the model parameters. The attack procedure in each frame follows four sequential phases.
\begin{enumerate}
    \setlength{\itemsep}{0pt}
    \setlength{\parskip}{0pt}
    \setlength{\parsep}{0pt}
    \item \textit{Local Perception Phase}: All agents, including the malicious one, process their sensing data independently and extract intermediate features using feature encoders.
    \begin{equation}
    \mathbf{F}_k = f_\mathtt{encoder}(\mathbf{O}_k), \quad k \in \mathcal{X}^N
    \end{equation}
    This phase operates in parallel without inter-agent communication.

    \item \textit{Feature Communication Phase}: All agents broadcast their extracted features through the network. Malicious agent $k$ collects feature information $\{\mathbf{F}_{j\rightarrow i}\}$ from other agents. Feature-level transmission ensures minimal communication overhead compared to raw sensor data exchange.

    \item \textit{Attack Generation Phase}: A malicious agent executes the attack by first perturbing its local features and then propagating them through the collaborative perception pipeline described in Section \ref{sec:formulation}.
    The attacker aims to optimize the perturbation $\delta$ through an iterative process. The optimization objective is formulated as:
    \begin{equation}
        \vspace{-2mm}
        \begin{aligned}
        \mathop{\arg\max}_{\delta} \mathcal{L}(\mathbf{Y}^\delta, \mathbf{Y}^\mathtt{gt}),
        \quad \mathtt{s.t.}\quad  \|\delta\|\leq \Delta
        \end{aligned}
    \end{equation}
    where $\Delta$ bounds the perturbation magnitude to maintain attack stealthiness. The total loss function is designed to aggregate adversarial losses over all object proposals, targeting both classification and localization aspects:
    \begin{equation}
        \mathcal{L}(\mathbf{Y}^\delta, \mathbf{Y}^\mathtt{gt}) = \sum_{p \in \mathbf{Y}^\delta} \mathcal{L}_\mathtt{adv}(p, p^\mathtt{gt})
    \end{equation}
    For each proposal $p$ with the highest confidence class $c = \mathop{\arg\max}\{p_i\}$, we leverage a class-specific adversarial loss following \citep{tuAdversarialAttacksMultiAgent2021}:
    \begin{equation*}
        \mathcal{L}_\mathtt{adv}(p', p) = \begin{cases}
            -\log(1 - p'_c)\cdot\eta & c \neq k,\ p_c > \tau_1\\
            -\lambda p'_c\log(1 - p'_c) & c = k,\ p_c > \tau_2\\
            0 & \text{otherwise}
        \end{cases}
    \end{equation*}
    where $\eta$ represents the IoU between perturbed and original proposals to consider spatial accuracy, $\tau_1$ and $\tau_2$ are confidence thresholds for different attack scenarios, $\lambda$ balances the importance of different attack objectives, and $k$ denotes the background class.

    \item \textit{Defense and Final Perception Phase}: The ego vehicle integrates all received feature information, including potentially corrupted ones, to complete the final object detection task. Note that we focus exclusively on CP-specific vulnerabilities, excluding physical sensor attacks (e.g., LiDAR or GPS spoofing), which are general threats to CAVs. We also assume communication channels are secured with proper cryptographic protection.
\end{enumerate}
The attack process is shown in Algorithm \ref{alg:attack_process}.

\section{CP-GuardBench}
\label{sec:cp_guardbench}

To facilitate feature-level malicious agent detection in CP systems, we present CP-GuardBench, the first dataset for malicious agent detection in CP systems. It provides a comprehensive dataset for training and evaluating malicious agent detection methods.
In this section, we will introduce the details of CP-GuardBench, including the automatic data generation and annotation pipeline in Section \ref{sec:data_generation}, and the data visualization and statistics in Section \ref{sec:data_statistics}.

\subsection{Automatic Data Generation and Annotation}
\label{sec:data_generation}

\begin{figure*}[t]
    \centering

    \includegraphics[width=.9\linewidth]{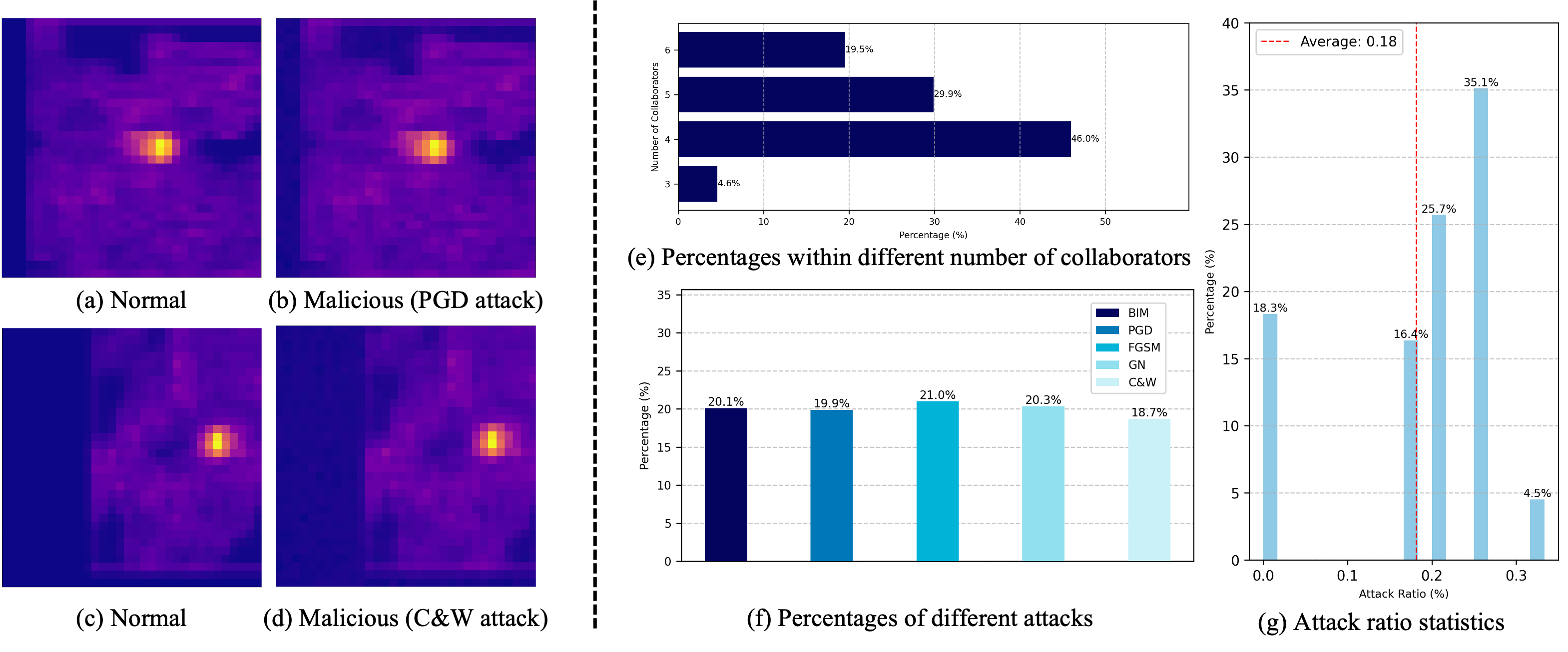}
    \caption{\textbf{Visualization and Statistics of CP-GuardBench. (a), (b), (c)  and (d) are visualization}, which visualizes the normal intermediate features and the adversarial examples perturbed by different malicious agents. We can see the adversarial examples are almost identical to the normal examples, which indicates the challenges in detecting malicious agents. \textbf{(e), (f), (g) and (h) are the statistics} of CP-GuardBench, including the number of collaborators, attack ratio, and attack types.}
    \label{fig:data_statistics}
    \vspace{-5mm}
\end{figure*}

We build CP-GuardBench based on one of the most widely used datasets in the CP field, V2X-Sim \citep{liV2XSimMultiAgentCollaborative2022}, which is a comprehensive simulated multi-agent perception dataset for V2X-aided autonomous driving.
In this section, we introduce the automatic data generation and annotation pipeline of CP-GuardBench. The pipeline is shown in Figure \ref{fig:data_generation}. It consists of three steps: 1) intermediate feature generation, 2) attack implementation, and 3) pair generation and saving.

Specifically, we first train a robust LiDAR collaborative object detector, which consists of a convolutional backbone, a convolutional decoder, and a prediction head for classification and regression \citep{Luo_2018_CVPR}. As for the fusion method, we adopt the mean fusion method to fuse the intermediate features from different collaborators.
Subsequently, the backbone is retained for extracting intermediate features, which are then transmitted and utilized by an ego CAV as supplementary information.

Secondly, the attacks are implemented and applied to the intermediate features.
The detection head and decoder are then frozen to generate the attacked detection results and optimize the adversarial perturbations.
As shown in Figure \ref{fig:data_generation}, several iterations are required to optimize the perturbations, and the loss function differs for different attack types.
In our CP-GuardBench, we consider five types of attacks, including Projected Gradient Descent (PGD) \citep{madry2018towards}, Carlini \& Wagner (C\&W) attack \citep{carlini2017evaluatingrobustnessneuralnetworks}, Basic Iterative Method (BIM) \citep{kurakin2017adversarialexamplesphysicalworld}, Gaussian Noise Perturbation (GN), and Fast Gradient Sign Method (FGSM) \citep{goodfellow2015explainingharnessingadversarialexamples}.
The details of these attacks are as follows:
\subsubsection{Projected Gradient Descent (PGD)} The mathematical formulation for PGD is as follows:
\begin{align}
    &\mathbf{F}_k^0 = \mathbf{F}_k + \text{Uniform}(-\Delta, \Delta) \\
    &\mathbf{F}_k^{t+1} = \Pi_{\Delta}\{\mathbf{F}_k^t + \alpha \cdot \text{sign}(\nabla_{\mathbf{F}_k^t} \mathcal{L}(\mathbf{F}_k^t, \mathbf{y}))\}
\end{align}
where $t$ is the iteration index, $\alpha$ is the step size, $\Delta$ is the maximum perturbation allowed, $\Pi_{\Delta}$ is the projection operation that ensures the perturbation is within the $\Delta$-ball of $\mathbf{F}_k$. The process is repeated for a fixed number of iterations $T$. In our implementation, we set $T=15$ and $\alpha = 0.1$.

\subsubsection{Carlini \& Wagner (C\&W) attack}
The C\&W attack aims to find the smallest perturbation $\delta$ that can cause misclassification. It can be formulated as an optimization problem:
\begin{equation}
    \min_{\delta} \|\delta\|_p + c \cdot f(\mathbf{F}_k + \delta)
\end{equation}
where $\|\cdot\|_p$ is the $L_p$ norm, $c > 0$ is a constant, and $f$ is an objective function that encourages misclassification:
\begin{equation}
    f(\mathbf{F}_k') = \max(\max_{i \neq t} Z(\mathbf{F}_k')_i - Z(\mathbf{F}_k')_t, -\kappa)
\end{equation}
Here, $Z(\mathbf{F}_k')$ is the logit output of the model, $t$ is the target class, and $\kappa$ is a confidence parameter.

\subsubsection{Basic Iterative Method (BIM)} BIM is similar to PGD but without the random initialization step. The mathematical formulation for BIM is as follows:
\begin{equation}
    \mathbf{F}_k^{t+1} = \text{Clip}_{\Delta}\{\mathbf{F}_k^t + \alpha \cdot \text{sign}(\nabla_{\mathbf{F}_k^t} \mathcal{L}(\mathbf{F}_k^t, \mathbf{y}))\}
\end{equation}
where $t$ is the iteration index, $\alpha$ is the step size, $\Delta$ is the maximum perturbation allowed, $\text{Clip}_{\Delta}$ is a function that clips the values to be within the $\Delta$-neighborhood of the original features $\mathbf{F}_k$, and $\mathbf{F}_k^0 = \mathbf{F}_k$. The process is repeated for a fixed number of iterations or until a stopping criterion is met.

\subsubsection{Fast Gradient Sign Method (FGSM)}
FGSM generates adversarial examples by perturbing the input in the direction of the gradient of the loss function with respect to the input. Given the intermediate features $\mathbf{F}_k$, the mathematical formulation of FGSM is as follows:
\begin{equation}
    \mathbf{F}_k^\text{adv} = \mathbf{F}_k + \Delta \cdot \text{sign}(\nabla_{\mathbf{F}_k} \mathcal{L}(\mathbf{F}_k, \mathbf{y}))
\end{equation}
where $\mathbf{F}_k^\text{adv}$ is the adversarial example, $\Delta$ is the perturbation magnitude, $\mathcal{L}$ is the loss function, and $\mathbf{y}$ is the true label. The $\text{sign}(\cdot)$ function takes the sign of the gradient, ensuring that the perturbation is in the direction that maximizes the loss.

\subsubsection{Gaussian Noise Perturbation (GN)}
This attack suits the scenario where an attacker has no information about the victim's model. It means that the attacker can only launch black-box attacks. In this attack, the attacker generates Gaussian noise $\delta_\mathtt{GN}$ and perturbs the original features $\mathbf{F}_k$.

In the generation of attack data, we randomly choose one of the attacks mentioned above and generate the corresponding attack data in each iteration.
Finally, the perturbed features will be annotated with the corresponding attack type and saved for later use.

\subsection{Data Visualization and Statistics}
\label{sec:data_statistics}

We visualize the samples of the generated data in Figures \ref{fig:data_statistics}(a), (b), (c), and (d). We observe that attacks are so stealthy that it is very hard to see the difference with the naked eye, which poses a great challenge to address the detection of malicious agents.

To construct CP-GuardBench, we randomly sample 9000 frames from V2X-Sim and generate 42200 feature-label pairs. The data is then split into training, validation, and test sets with a ratio of 8:1:1.
The data statistics are shown in Figures \ref{fig:data_statistics}(e), (f), and (g). Figure \ref{fig:data_statistics}(e) illustrates the distribution of the number of collaborators, which is the number of agents that collaboratively perceive environments.
The number of collaborators ranges from 3 to 6, with the most common scenario being 4 collaborators, accounting for 46.0\% of the total data. 5 and 6 collaborators are also common, accounting for 29.9\% and 19.5\% of the total data, respectively.
Regarding the distribution of attack types, as depicted in Figure \ref{fig:data_statistics}(f), we observe that the attack types are evenly distributed, with each type accounting for approximately 20\% of the total data. This is due to the random selection of one attack type in each iteration. Figure \ref{fig:data_statistics}(g) illustrates the attack ratio, which represents the ratio of the number of attackers to the total number of agents in a collaboration network. The maximum attack ratio exceeds 0.3, the minimum is 0, and the average attack ratio is 0.18.

\subsection{Discussion of the Benefits of CP-GuardBench}

In this section, we discuss the benefits and the contribution of CP-GuardBench. The reason why we develop CP-GuardBench is to facilitate the training and evaluation of feature-level defense methods for CP systems against general adversarial attacks. Since this work is the first attempt to leverage feature-level knowledge for CP defense, CP-GuardBench is the first dataset in this field. We hope this dataset will help the community to develop more effective feature-level defense methods for CP systems.

\section{CP-Guard+}
\label{sec:cp_guard+}

In this section, we present our CP-Guard+, a tailored defense method for CP scenarios that effectively detects malicious agents. It consists of two techniques: 1) Residual Latent Feature Learning, which learns the residual features of benign and malicious agents, and 2) Dual Centered Contrastive Loss (DCCLoss), which clusters the representation of benign features into a compact space and ensures that the representation of malicious features is as distant from the benign space as possible.

\subsection{Residual Latent Feature Learning}
In CP scenarios, the dynamic environment causes noisy, non-stationary data distributions. Directly detecting malicious agents can be challenging due to this noise, as object detectors' feature maps often mix foreground and background information.

To tackle this challenge, we propose to learn residual latent features instead of directly learning the features of benign or malicious agents. By focusing on the differences between the collaborators' feature maps and the ego agent's feature maps, the model can better distinguish between benign and malicious agents. This mechanism is also inspired by the idea that the collaborators' intermediate feature maps will achieve a consensus rather than conflict with the ego CAV's intermediate feature maps.

Specifically, given the collaborators' intermediate feature maps $\{\mathbf{F}_{j\rightarrow i}\}_{j\neq i,\  j\in \mathcal{X}^N}$ and the ego CAV's intermediate feature maps $\mathbf{F}_i$, the residual feature is obtained as: $\mathbf{F}^\mathtt{res}_{j\rightarrow i}= \mathbf{F}_i - \mathbf{F}_{j\rightarrow i}.$
We adopt subtraction rather than concatenation of $(\mathbf{F}_i, \mathbf{F}_{j\rightarrow i})$ for two reasons. First, the residual $\mathbf{F}_i - \mathbf{F}_{j\rightarrow i}$ isolates inter-agent disagreement from shared background content: benign pairs yield near-zero residuals while malicious perturbations produce large, discriminative conflicts. Second, subtraction preserves the original feature dimensionality, avoiding the parameter overhead and latency of concatenation—a critical advantage for real-time collaborative perception.
Then, we can leverage the residual latent features to detect malicious agents by modeling the detection problem as a binary classification task. A binary classifier $f_{\mathtt{cls}}(\mathbf{x}; \theta)$ is trained on the residual latent features to distinguish between benign (labeled 0) and malicious (labeled 1) agents. The model is optimized using the cross-entropy loss $\mathcal{L}_\mathtt{CE}$:
\begin{equation}
    \min_\theta \mathcal{L}_\mathtt{CE} (f_{\mathtt{cls}}(\mathbf{F}^\mathtt{res}_{j\rightarrow i}; \theta), y_{j\rightarrow i})
\end{equation}
where $y_{j\rightarrow i}$ is the ground truth label of residual feature $\mathbf{F}^\mathtt{res}_{j\rightarrow i}$.

\begin{table*}[t]
    \caption{\textbf{Performance Evaluation of CP-Guard+ on CP-GuardBench.} We report the average accuracy, true positive rate (TPR), false positive rate (FPR), precision, and F1 score of CP-Guard+ on CP-GuardBench with different attack methods and perturbation budgets $\Delta = 0.10, 0.25, 0.5, 0.75$. }
    \label{tab:quantitative_results_cpguardbench}
    \centering

    \renewcommand\arraystretch{1.0}
    \resizebox{1\linewidth}{!}{
    \small
    \begin{tabular}{p{1.2cm}|ccccc|ccccc}
        \toprule

        \multirow{2}{*}{ \textbf{Metrics} } & \multicolumn{5}{c|}{$\Delta = 0.10$} & \multicolumn{5}{c}{$\Delta = 0.25$} \\

        & Accuracy $\uparrow$ & TPR $\uparrow$  & FPR $\downarrow$ & Precision $\uparrow$ & F1 Score $\uparrow$ & Accuracy $\uparrow$ & TPR $\uparrow$ & FPR $\downarrow$ & Precision $\uparrow$ & F1 Score $\uparrow$ \\ \midrule

        PGD&99.79&98.97&0.00&100.00&99.48&99.93&100.00&0.09&99.66&99.83\\
        BIM&99.93&100.00&0.09&99.66&99.83&100.00&100.00&0.00&100.00&100.00\\
        C\&W &99.73&98.96&0.09&99.65&99.30&99.86&100.00&0.17&99.32&99.66\\

        FGSM&91.23&56.80&0.09&99.40&72.29&98.77&94.24&0.09&99.64&96.86\\
        GN &91.02&55.63&0.09&99.39&71.33&98.49&92.88&0.09&99.64&96.14\\ \midrule
        Average &96.34&82.07&0.07&99.62&88.45&99.41&97.42&0.09&99.65&98.50\\ \bottomrule\toprule

        \multirow{2}{*}{ \textbf{Metrics} } & \multicolumn{5}{c|}{$\Delta = 0.50$} & \multicolumn{5}{c}{$\Delta = 0.75$} \\

        & Accuracy $\uparrow$ & TPR $\uparrow$ & FPR $\downarrow$  & Precision $\uparrow$ & F1 Score $\uparrow$ & Accuracy $\uparrow$ & TPR $\uparrow$ & FPR $\downarrow$ & Precision $\uparrow$ & F1 Score $\uparrow$ \\ \midrule

        PGD&99.86&100.00&0.17&99.32&99.66&99.86&100.00&0.17&99.33&99.66\\
        BIM&99.93&100.00&0.09&99.66&99.83&99.93&100.00&0.09&99.66&99.83\\
        C\&W &99.86&100.00&0.17&99.32&99.66&99.93&100.00&0.09&99.66&99.83\\
        FGSM&99.93&100.00&0.09&99.66&99.83&99.79&99.66&0.17&99.32&99.49\\
        GN &99.86&100.00&0.17&99.32&99.66&99.79&99.65&0.17&99.31&99.48\\ \midrule
        Average &99.89&100.00&0.14&99.46&99.73&99.86&99.86&0.14&99.46&99.66\\ \bottomrule
    \end{tabular}
    }
\end{table*}

\begin{table*}[t]
    \caption{\textbf{Comparative results of CP-Guard+ on the V2X-Sim Dataset.} We report AP@0.5 and AP@0.7 with different perturbation budgets $\Delta$ and the number of malicious agents $N_\text{mal}$. }
    \label{tab:quantitative_results_v2xsim}
    \centering

    \renewcommand\arraystretch{0.9}
    \resizebox{1\linewidth}{!}{
    \small
    \begin{tabular}{c|c|cc|cc|cc|cc}
        \toprule

        \multirow{2}{*}{ {\makecell{Attack\\Method}} }&\multirow{2}{*}{ {\makecell{Defense\\Method}} } & \multicolumn{2}{c|}{$\Delta = 0.25, N_\text{mal} = 1$ } & \multicolumn{2}{c|}{$\Delta = 0.5, N_\text{mal} = 1$} & \multicolumn{2}{c|}{$\Delta = 0.25, N_\text{mal} = 2$} & \multicolumn{2}{c}{$\Delta = 0.5, N_\text{mal} = 2$}\\
               && AP@0.5 & AP@0.7  & AP@0.5 & AP@0.7 & AP@0.5 & AP@0.7  & AP@0.5 & AP@0.7 \\
        \midrule
        \multirow{4}{*}{PGD } & No Defense & 29.73 & 28.47  & 11.35 & 11.17 & 12.69 & 12.42 & 1.69 & 1.65 \\
        & MADE & 64.63 & 45.22 & 64.81 & 44.89 & 62.45 & 43.49 & 63.04 & 43.77 \\
        & ROBOSAC & 62.13 & 42.90 & 63.67 & 43.79 & 59.01 & 40.03 & 59.97 & 40.44 \\
        & CP-Guard+ & \textbf{72.89} & \textbf{71.45} & \textbf{69.50} & \textbf{68.56} & \textbf{69.50} & \textbf{67.92} & \textbf{66.09} & \textbf{64.82} \\
        \midrule

        \multirow{4}{*}{C\&W }
        & No Defense & 19.03 & 16.58 & 4.69 & 3.78 & 19.03 & 16.58 & 0.71 & 0.58 \\
        & MADE & 65.26 & 45.24 & \textbf{64.74} & 45.65 & 63.41 & 44.28  & \textbf{62.86} & 42.93 \\
        & ROBOSAC & 61.83 & 42.01 & 62.47 & 42.80 & 59.39 & 39.94  & 59.83 & 39.82 \\
        & CP-Guard+ & \textbf{69.41} & \textbf{66.86} & {60.64} & \textbf{55.41} & \textbf{64.17} & \textbf{61.73}  & {58.54} & \textbf{53.15} \\
        \midrule

        \multirow{4}{*}{BIM }
        & No Defense & 26.69 & 25.71 & 10.05 & 9.89 & 11.59 & 11.38 & 1.37 & 1.33 \\
        & MADE & 66.11 & 45.94 & 65.51 & 45.47 &  64.36 & 43.89  & 63.56 & 44.09 \\
        & ROBOSAC & 62.69 & 43.80  & 63.78 & 43.66 & 59.10 & 39.74  & 59.29 & 39.89 \\
        & CP-Guard+ & \textbf{73.35} & \textbf{71.46} & \textbf{66.83} & \textbf{66.05} & \textbf{70.91} & \textbf{69.11}  & \textbf{66.30} & \textbf{64.62} \\
        \midrule

        \multirow{4}{*}{Average}
        & No Defense & 25.15 & 23.59 & 8.70 & 8.28 & 14.44 & 13.46 & 1.27 & 1.19 \\
        & MADE & 65.33 & 45.47 & 65.02 & 45.34 & 63.41 & 43.89 & 63.15 & 43.60 \\
        & ROBOSAC & 62.21 & 42.90 & 63.31 & 43.42 & 59.37 & 39.90 & 59.70 & 40.05 \\
        & CP-Guard+ & \textbf{71.88} & \textbf{69.92} & \textbf{65.66} & \textbf{63.34} & \textbf{68.19} & \textbf{66.25} & \textbf{63.64} & \textbf{60.86}\\
        \midrule
        \multicolumn{2}{c|}{Upper-bound} & 79.94 & 78.40& 79.94 & 78.40 & 79.94 & 78.40 & 79.94 & 78.40 \\
        \bottomrule
    \end{tabular}
    }
    \vspace{-5mm}
\end{table*}

\subsection{Dual-Centered Contrastive Loss}

In practice, attackers can continuously design new attacks to manipulate the victim's intermediate feature maps. Therefore, we need a model that is resistant to unseen attacks, which means the model should cluster the representation of benign features into a more compact space and ensure that the representation of malicious features is as distant from the benign space as possible.

To tackle this challenge, we propose a Dual-Centered Contrastive Loss (DCCLoss), which is a contrastive learning-based objective function. It explicitly models the distribution relationship between benign and malicious features, pulling positive pairs (features of the same class) closer to their corresponding center, thereby enhancing the internal consistency of both benign and malicious features. Meanwhile, negative pairs (features from different classes) will be pushed away from each other's centers, ensuring maximal separation between benign and malicious features in the feature space. In this way, the robustness of the model against unseen attacks is improved.

Specifically, we first leverage the output of the penultimate fully connected layer $f_{\text{cls}}$ to obtain the one-dimensional vector $\{\mathcal{V}_i\}_{i=0,1,\ldots, N-1}$ of residual features $\{\mathbf{F}^\mathtt{res}_{j\rightarrow i}\}_{j\neq i,\  j\in \mathcal{X}^N}$. Then, we introduce two feature centers in DCCLoss:
\begin{enumerate}
    \item Benign feature center ($\mathbf{c}_{\mathtt{b}}$): represents the center of all benign features, ensuring a compact distribution of benign features.
    \item Malicious feature center ($\mathbf{c}_{\mathtt{mal}}$): represents the center of all malicious features, ensuring maximal separation between malicious and benign feature distributions.
\end{enumerate}
The benign feature center $\mathbf{c}_\mathtt{b}$ and the malicious feature center $\mathbf{c}_\mathtt{mal}$ are computed by averaging the vectors of benign and malicious features, respectively:
\begin{equation}
    \mathbf{c}_\mathtt{b} = \frac{1}{N_\mathtt{b}}\sum_{\mathcal{V}_i\in \{\mathcal{V}_\mathtt{b}\}}^{N_\mathtt{b}} \mathcal{V}_i, \quad  \mathbf{c}_\mathtt{mal} = \frac{1}{N_\mathtt{mal}}\sum_{\mathcal{V}_i\in \{\mathcal{V}_\mathtt{mal}\}}^{N_\mathtt{mal}} \mathcal{V}_i.
\end{equation}
where $N_\mathtt{b}$ and $N_\mathtt{mal}$ are the numbers of benign and malicious vectors, respectively, and $\mathcal{V}_\mathtt{b}$ and $\mathcal{V}_\mathtt{mal}$ are the sets of benign and malicious vectors, respectively.

Moreover, denote $(\mathcal{V}_m, \mathcal{V}_n)$ as a pair of features, which is a positive pair if they are from the same class (both benign or malicious) and a negative pair otherwise. The DCCLoss for a single feature pair $\ell(\mathcal{V}_m, \mathcal{V}_n)$ is defined as:
\begin{equation}
    \begin{aligned}
    &\ell\left(\mathcal{V}_m, \mathcal{V}_n\right)=-\log \\
    &\frac{\exp\left((\mathcal{V}_m-\mathbf{c}^{(m)})\odot  (\mathcal{V}_n-\mathbf{c}^{(n)})/\tau\right)}
        {\sum_{o=1, o\neq m}^{N} \mathbb{I}\cdot\exp\left((\mathcal{V}_m-\mathbf{c}^{(m)})\odot  (\mathcal{V}_o-\mathbf{c}^{(o)})/\tau\right)},\\
    & \text{where}\ \mathcal{V}_x - \mathbf{c}^{(x)}=
    \left\{\begin{aligned}
        &\mathcal{V}_x - \mathbf{c}_\mathtt{b}, & \text{if } \mathcal{V}_x \in \{\mathcal{V}_\mathtt{b}\},\\
        &\mathcal{V}_x - \mathbf{c}_\mathtt{mal}, & \text{if } \mathcal{V}_x \in \{\mathcal{V}_\mathtt{mal}\},
    \end{aligned}\right.
    \end{aligned}
\end{equation}
where $\mathbb{I}(\mathcal{V}_m, \mathcal{V}_o)$ is an indicator function that returns one or zero for positive or negative pairs, respectively. $\tau$ is a temperature parameter and $\odot$ denotes the cosine similarity, where $\mathcal{V}_m\odot \mathcal{V}_n = \frac{\mathcal{V}_m^\top  \mathcal{V}_n}{\|\mathcal{V}_m\|\|\mathcal{V}_n\|}$.
The final DCCLoss is the average of $\ell$ of all positive pairs.
\begin{equation}
    \mathcal{L}_\mathtt{DCCLoss}= \sum_{m=1}^{N} \sum_{n=m+1}^{N} \frac{\left(1-\mathbb{I}(\mathcal{V}_m, \mathcal{V}_n)\right)\ell\left(\mathcal{V}_m, \mathcal{V}_n\right)}{C(N,2)},
    \label{eq:contrastive_loss}
\end{equation}
where $C(N,2)= \binom{N}{2}={N!}/({2!(N-2)!})$.
During training, we use the combination of cross entropy loss and Eq. \ref{eq:contrastive_loss} to optimize the model:
\begin{equation}
    \mathcal{L} = \mathcal{L}_\mathtt{CE} + \alpha\cdot \mathcal{L}_\mathtt{DCCLoss}
    \label{eq:mixed_loss}
\end{equation}
where $\alpha$ is a hyperparameter to balance the two losses.

\textbf{Discussion of DCCLoss.} Accordingly, the first term $\mathcal{L}_\mathtt{CE}$ quantifies the difference between the true distribution and the predicted distribution from the model, thereby penalizing the confidence in wrong predictions. The second term $\mathcal{L}_\mathtt{DCCLoss}$ contributes significantly to the learning process. Standard contrastive loss attempts to maximize the distance between negative pairs, which may cause the features of benign samples to gradually drift away from their center. However, DCCLoss calculates the distance using the feature center as a reference point, thus avoiding this issue. The optimization goal of DCCLoss is to maximize the angular distance between negative pairs to enhance feature discriminability while maintaining the compactness of benign sample features, keeping them as close to the feature center as possible.
In other words, the introduction of dual-center modeling optimizes the distributional relationship between benign and malicious features, making them more separable while simultaneously making each distribution more compact. This helps resolve the distribution overlap problem and enhances the model's ability to detect unseen attacks.

\begin{table*}[t]
    \caption{\textbf{Generalization Results of CP-Guard+.} We leverage the leave-one-out strategy to evaluate the generalization ability of CP-Guard+. We report the average accuracy, true positive rate (TPR), false positive rate (FPR), precision, and F1 score of CP-Guard+ on CP-GuardBench with different attack methods and perturbation budgets $\Delta = 0.10, 0.25, 0.5, 0.75$. }
    \vspace{-2mm}
    \label{tab:generalization_results}
    \centering

    \renewcommand\arraystretch{1.0}
    \resizebox{1\linewidth}{!}{
    \small
    \begin{tabular}{c|ccccc|ccccc}
        \toprule

        \multirow{2}{*}{ {Held-out} } & \multicolumn{5}{c|}{$\Delta = 0.10$} & \multicolumn{5}{c}{$\Delta = 0.25$} \\

        & Accuracy $\uparrow$ & TPR $\uparrow$  & FPR $\downarrow$ & Precision $\uparrow$ & F1 Score $\uparrow$ & Accuracy $\uparrow$ & TPR $\uparrow$ & FPR $\downarrow$ & Precision $\uparrow$ & F1 Score $\uparrow$ \\ \midrule

        PGD&98.83&94.48&0.09&99.64&96.99&99.93&100.00&0.09&99.66&99.83\\
        BIM&99.25&100.00&0.94&96.38&98.16&100.00&100.00&0.00&100.00&100.00\\
        C\&W &97.05&85.76&0.17&99.20&91.99&99.86&100.00&0.17&99.32&99.66\\

        FGSM&85.61&28.57&0.00&100.00&44.44&98.77&94.24&0.09&99.64&96.86\\
        GN &91.64&59.04&0.17&98.86&73.93&98.49&92.88&0.09&99.64&96.14\\ \midrule
        Average &94.48&73.57&0.27&98.82&81.10&98.23 & 92.94 & 0.22 & 99.08 & 95.20\\ \midrule
        Upper-bound &96.34&82.07&0.07&99.62&88.45&99.41&97.42&0.09&99.65&98.50\\ \bottomrule\toprule

        \multirow{2}{*}{ {Held-out} } & \multicolumn{5}{c|}{$\Delta = 0.50$} & \multicolumn{5}{c}{$\Delta = 0.75$} \\

        & Accuracy $\uparrow$ & TPR $\uparrow$ & FPR $\downarrow$  & Precision $\uparrow$ & F1 Score $\uparrow$ & Accuracy $\uparrow$ & TPR $\uparrow$ & FPR $\downarrow$ & Precision $\uparrow$ & F1 Score $\uparrow$ \\ \midrule

        PGD&99.93&100.00&0.09&99.66&99.83&99.93&100.00&0.09&99.66&99.83\\
        BIM&99.31&100.00&0.86&96.68&98.31&99.25&100.00&0.94&96.37&98.15\\
        C\&W &99.93&100.00&0.09&99.66&99.83&99.86&100.00&0.17&99.33&99.66\\
        FGSM&99.93&100.00&0.09&99.66&99.83&99.93&100.00&0.09&99.66&99.83\\
        GN &99.79&100.00&0.26&98.98&99.49&99.86&100.00&0.17&99.31&99.66\\ \midrule
        Average &99.78&100.00&0.28&98.93&99.66&99.77&100.00&0.29&98.87&99.43\\ \midrule
        Upper-bound &99.89&100.00&0.14&99.46&99.73&99.86&99.86&0.14&99.46&99.66\\ \bottomrule
    \end{tabular}
    }
    \vspace{-5mm}
\end{table*}

\section{Experiments}
\label{sec:experiments}

\subsection{Experimental Setup}

\subsubsection{Dataset and Baselines} In our experiments, we consider two datasets: CP-GuardBench and V2X-Sim \citep{liV2XSimMultiAgentCollaborative2022}.
We designate CAV \#1 as the ego CAV and randomly select adversarial collaborators from the remaining CAVs. Additionally, we use ROBOSAC \citep{liUsAdversariallyRobust2023} and MADE \citep{zhaoMaliciousAgentDetection2024} as baselines, which are two state-of-the-art CP defense methods based on the hypothesize-and-verify paradigm.

\subsubsection{Attack Settings} We assess different CP defense methods targeted at five attacks: PGD attack \citep{madry2018towards}, C\&W attack \citep{carlini2017evaluatingrobustnessneuralnetworks}, BIM attack \citep{kurakin2017adversarialexamplesphysicalworld}, FGSM attack \citep{goodfellow2015explainingharnessingadversarialexamples}, and GN attack. We set different perturbation sizes $\Delta \in \{0.1, 0.25, 0.5, 0.75, 1.0\}$. The number of malicious attackers varies in $\{0,1,2\}$ and all the attackers are randomly assigned from the collaborators, where 0 attacker indicates an upper-bound case. For PGD, BIM and C\&W attacks, the number of iteration steps is 15 and the step size is 0.1.

\subsubsection{Implementation Details} The CP-Guard+ system is implemented using PyTorch, and we utilize the object detector described in Section \ref{sec:data_generation}.
For each agent, the local LiDAR point cloud data is first encoded into $32\times 32$ bird's eye view (BEV) feature maps with 256 channels prior to communication. For our CP-Guard+, we use ResNet-50 \citep{heDeepResidualLearning2016} as the backbone, and the training is performed for 50 epochs with batch size 10 and learning rate $1\times 10^{-3}$.  Our experiments are conducted on a server with four RTX A5000 GPUs.
For mixed contrastive training, we utilize the output of the fully connected layers preceding the final output layer in the backbone to form a one-dimensional feature vector for each agent, the dimension of which is 2048. The DCCLoss temperature is set to $\tau=0.07$ and the loss balancing weight to $\alpha=0.10$.

\subsubsection{Evaluation Metrics} We use a variety of metrics to evaluate the performance of our CP-Guard+ model. For malicious agent detection on our CP-GuardBench dataset, we consider Accuracy, True Positive Rate (TPR), False Positive Rate (FPR), Precision, and F1 Score. For malicious-agent detection on CP-GuardBench, a false positive refers to a benign agent that is incorrectly classified as malicious, and the false positive rate (FPR) is computed in the standard binary-classification manner. For CP defense on the V2X-Sim dataset, we use metrics including average precision (AP) at IoU=0.5 and IoU=0.7. Additionally, to assess the computation efficiency of different CP defense methods, we introduce the metric frames-per-second (FPS).

\subsection{Quantitative Results}

\subsubsection{Performance Evaluation of CP-Guard+}
Table \ref{tab:quantitative_results_cpguardbench} reports CP-Guard+ performance on CP-GuardBench across five attacks and four perturbation budgets. At $\Delta=0.25$, CP-Guard+ exceeds 98\% accuracy on PGD, BIM, and C\&W; FGSM and GN show lower accuracy at small $\Delta$ due to their weaker perturbations, but improve rapidly as $\Delta$ grows. Across all settings, CP-Guard+ maintains consistently high TPR and low FPR, demonstrating strong and robust detection.

\begin{figure*}[t]
    \centering
    \includegraphics[width=1\linewidth]{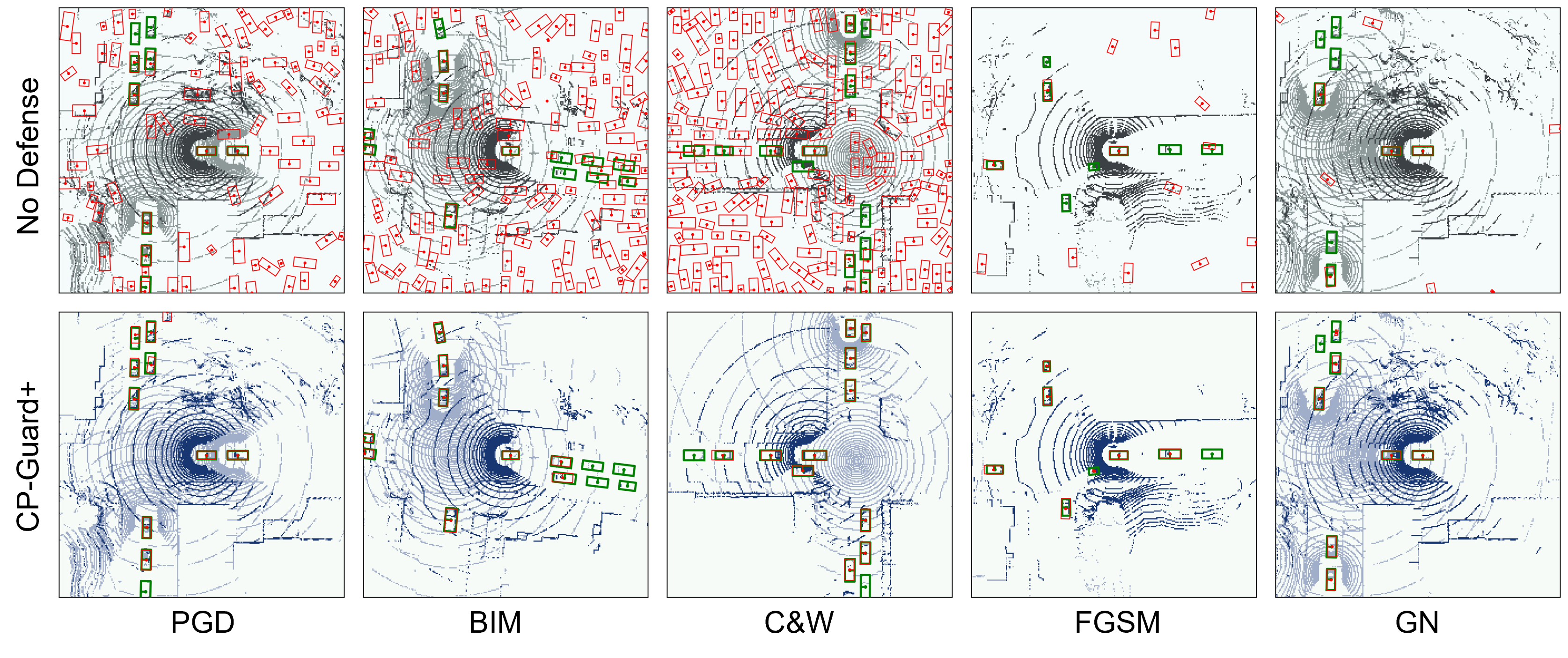}
    \vspace{-5mm}
    \caption{\textbf{Visualization and Qualitative Results.} We visualize the results of the CP systems with and without defense by CP-Guard+. The \textcolor{red}{red} bounding boxes represent the predicted outcomes, while the \textcolor{green}{green} ones denote the ground truth.}
    \label{fig:qualitative_results}
    \vspace{-5mm}
\end{figure*}

\begin{figure*}[t]
    \centering
    \includegraphics[width=.7\linewidth]{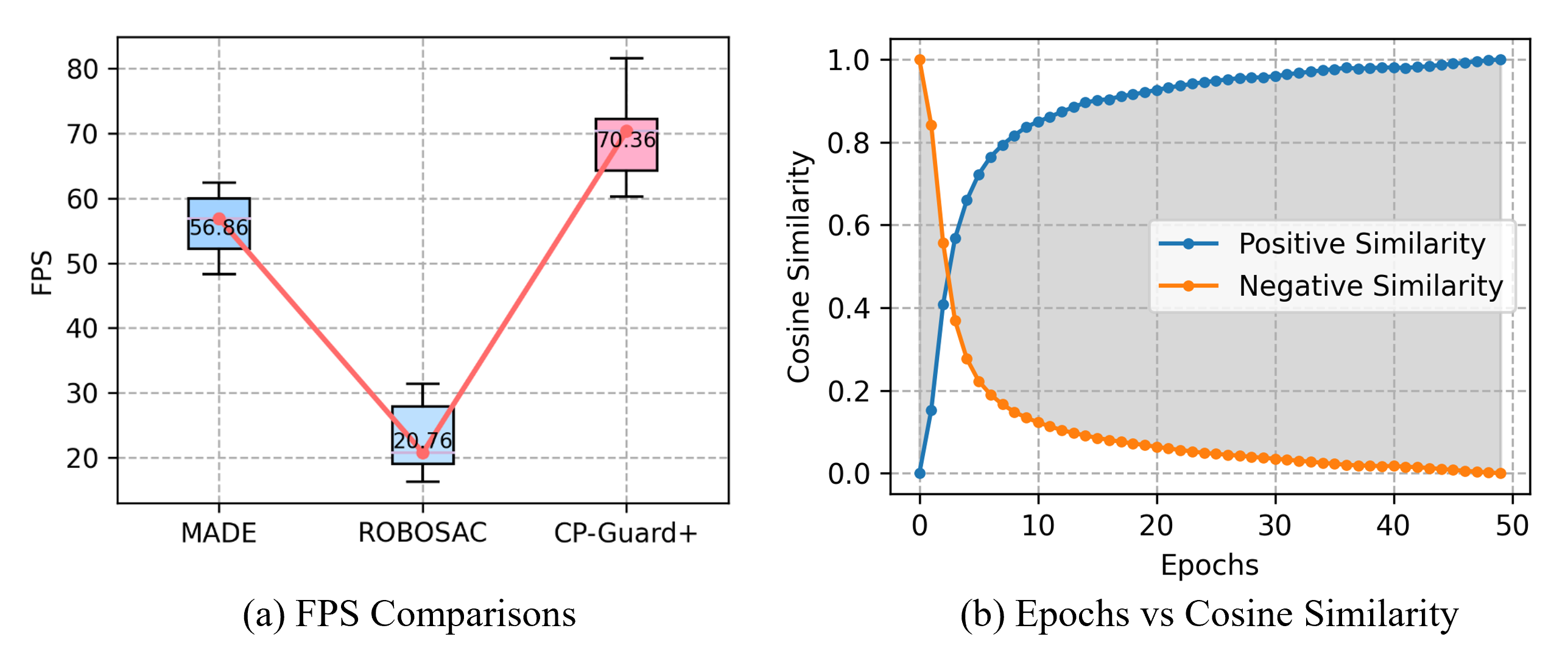}
    \caption{\textbf{(a) FPS performance comparison} between CP-Guard+ and other baselines. \textbf{(b) Cosine distance} between the intermediate features of the malicious agent and the benign agent on CP-GuardBench with $\Delta=0.25$.
    }
    \label{fig:cont_ablation}
    \vspace{-5mm}
\end{figure*}

\subsubsection{Performance Comparison with Other Defenses} We compare AP@0.5 and AP@0.7 of CP-Guard+ against MADE \citep{zhaoMaliciousAgentDetection2024} and ROBOSAC \citep{liUsAdversariallyRobust2023} on V2X-Sim (Table \ref{tab:quantitative_results_v2xsim}). Without defense, AP degrades sharply under attack; CP-Guard+ effectively recovers performance and consistently outperforms both baselines. At $\Delta=0.25$, $N_\text{mal}=1$, CP-Guard+ achieves 71.88\% AP@0.5 and 69.92\% AP@0.7, surpassing MADE by 10.03\%/53.77\% and ROBOSAC by 15.54\%/62.98\%. This advantage holds as $\Delta$ and $N_\text{mal}$ increase, demonstrating the robustness and superiority of CP-Guard+.

\begin{figure*}[t]
    \centering
    \includegraphics[width=.85\linewidth]{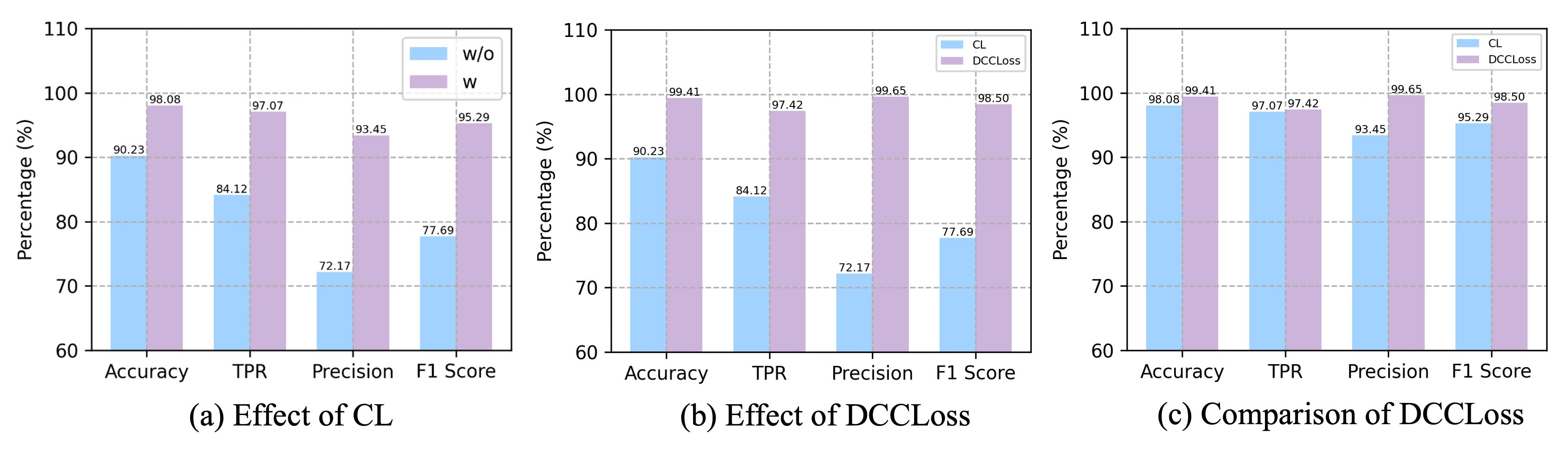}
    \caption{Evaluated on CP-GuardBench with $\Delta=0.25$. \textbf{(a) Effectiveness of contrastive loss (CL).} `w/o' means \textit{without} CL. `w/' means \textit{with} CL.  \textbf{(b) Effectiveness of DCCLoss.} `w/o' means \textit{without} DCCLoss. `w/' means \textit{with} DCCLoss. \textbf{(c) Comparison of DCCLoss.}
    }
    \label{fig:dccloss_ablation}
    \vspace{-5mm}
\end{figure*}

\subsubsection{FPS Comparison} We compare the FPS performance of CP-Guard+ with MADE and ROBOSAC, as shown in Figure \ref{fig:cont_ablation}(a). The median FPS values for MADE, ROBOSAC, and CP-Guard+ are 56.86, 20.76, and 70.36, respectively. CP-Guard+ achieves a 23.74\% higher FPS than MADE and a 238.92\% increase over ROBOSAC, representing a significant improvement. These results highlight the high computational efficiency of our CP-Guard+.

\subsubsection{Generalization of CP-Guard+} To evaluate the cross-attack generalization ability of our method, we conduct experiments using a leave-one-out strategy under the same underlying data distribution, where one attack type is excluded during training and used only for testing. In this approach, we iteratively excluded one type of attack from the training set, trained the model on the remaining attacks, and then tested its performance on the held-out attack type. The experimental results are presented in Table \ref{tab:generalization_results}. From the table, we can see that our method achieves strong generalization ability on unseen attacks. Compared to the upper-bound case, our method experiences only a marginal decrease in overall performance. These findings underscore our method's robust capability to detect and handle unseen attack patterns.

\subsection{Ablation Study and Visualization}

\subsubsection{The Effectiveness of DCCLoss}
We assess the impact of DCCLoss on CP-Guard+ performance. Figure \ref{fig:dccloss_ablation} shows the effectiveness of DCCLoss and its superiority compared with contrastive loss (CL) \cite{chenSimpleFrameworkContrastive2020}.
Figure \ref{fig:dccloss_ablation}(b) shows that this training significantly boosts performance, including Accuracy, TPR, Precision, and F1 score. Figure \ref{fig:dccloss_ablation}(c) shows the comparison results between DCCLoss and CL, which demonstrate that our DCCLoss outperforms CL and shows its effectiveness in malicious agent detection. In addition,
Figure \ref{fig:cont_ablation}(b) illustrates that DCCLoss increases the cosine distance between negative pairs and decreases it between positive pairs, enhancing the model's ability to distinguish between malicious and benign agents.

\subsubsection{Ablation Study on Residual Feature Formulation}
To examine the effect of the residual feature definition on detection sensitivity, we compare three feature interaction formulations at $\Delta=0.25$: subtraction ($\mathbf{F}_i - \mathbf{F}_{j\to i}$), concatenation ($[\mathbf{F}_i;\mathbf{F}_{j\to i}]$), and element-wise multiplication ($\mathbf{F}_i \odot \mathbf{F}_{j\to i}$). As shown in Table~\ref{tab:residual_ablation}, the residual formulation achieves the best performance across all metrics (99.41\% Accuracy, 97.42\% TPR, 99.65\% Precision, 98.50\% F1), confirming that subtraction most effectively suppresses the shared background content and highlights the inter-agent discrepancy patterns relevant to malicious-agent detection.

\begin{table*}[t]
    \centering
    \caption{\textbf{Ablation Study on Residual Feature Formulation} at $\Delta=0.25$.}
    \vspace{-2mm}
    \label{tab:residual_ablation}
    \renewcommand\arraystretch{1.0}
    \begin{tabular}{lcccc}
        \toprule
        {Formulation} & {Accuracy} & {TPR} & {Precision} & {F1} \\
        \midrule
        Concat ($[\mathbf{F}_i;\mathbf{F}_{j\to i}]$) & 92.21 & 62.03 & 95.12 & 63.80 \\
        Element-wise Mul ($\mathbf{F}_i \odot \mathbf{F}_{j\to i}$) & 96.92 & 87.03 & 97.38 & 91.54 \\
        Residual ($\mathbf{F}_i - \mathbf{F}_{j\to i}$) & \textbf{99.41} & \textbf{97.42} & \textbf{99.65} & \textbf{98.50} \\
        \bottomrule
    \end{tabular}
    \vspace{-2mm}
\end{table*}

\subsubsection{Ablation Study on DCCLoss Weight $\alpha$}
We further conduct an ablation study on the DCCLoss weight $\alpha$ at $\Delta=0.25$, testing $\alpha \in \{0.01, 0.05, 0.10, 0.20\}$. As shown in Table~\ref{tab:alpha_ablation}, CP-Guard+ remains strong across all tested values, indicating reasonable robustness to $\alpha$. The default $\alpha=0.10$ achieves the best overall Accuracy (99.41\%) and TPR (97.42\%), while other settings also maintain competitive performance (e.g., Accuracy above 97.93\% and F1 above 94.24\%).

\begin{table*}[t]
    \centering
    \caption{\textbf{Ablation Study on DCCLoss Weight $\alpha$} at $\Delta=0.25$.}
    \vspace{-2mm}
    \label{tab:alpha_ablation}
    \renewcommand\arraystretch{1.0}
    \begin{tabular}{lccccc}
        \toprule
        {$\alpha$} & {Accuracy} & {TPR} & {FPR} & {Precision} & {F1} \\
        \midrule
        0.05 & 97.93 & 90.44 & \textbf{0.17} & \textbf{99.22} & 94.24 \\
        0.01 & 98.45 & 94.78 & 0.62 & 97.46 & \textbf{96.00} \\
        0.20 & 98.41 & 93.69 & 0.39 & 98.35 & 95.80 \\
        0.10 (default) & \textbf{99.41} & \textbf{97.42} & {0.22} & {99.08} & {95.20} \\
        \bottomrule
    \end{tabular}
    \vspace{-3mm}
\end{table*}

\subsubsection{Feature Center Convergence During Training}
\begin{figure*}[t]
    \centering
    \includegraphics[width=0.36\linewidth]{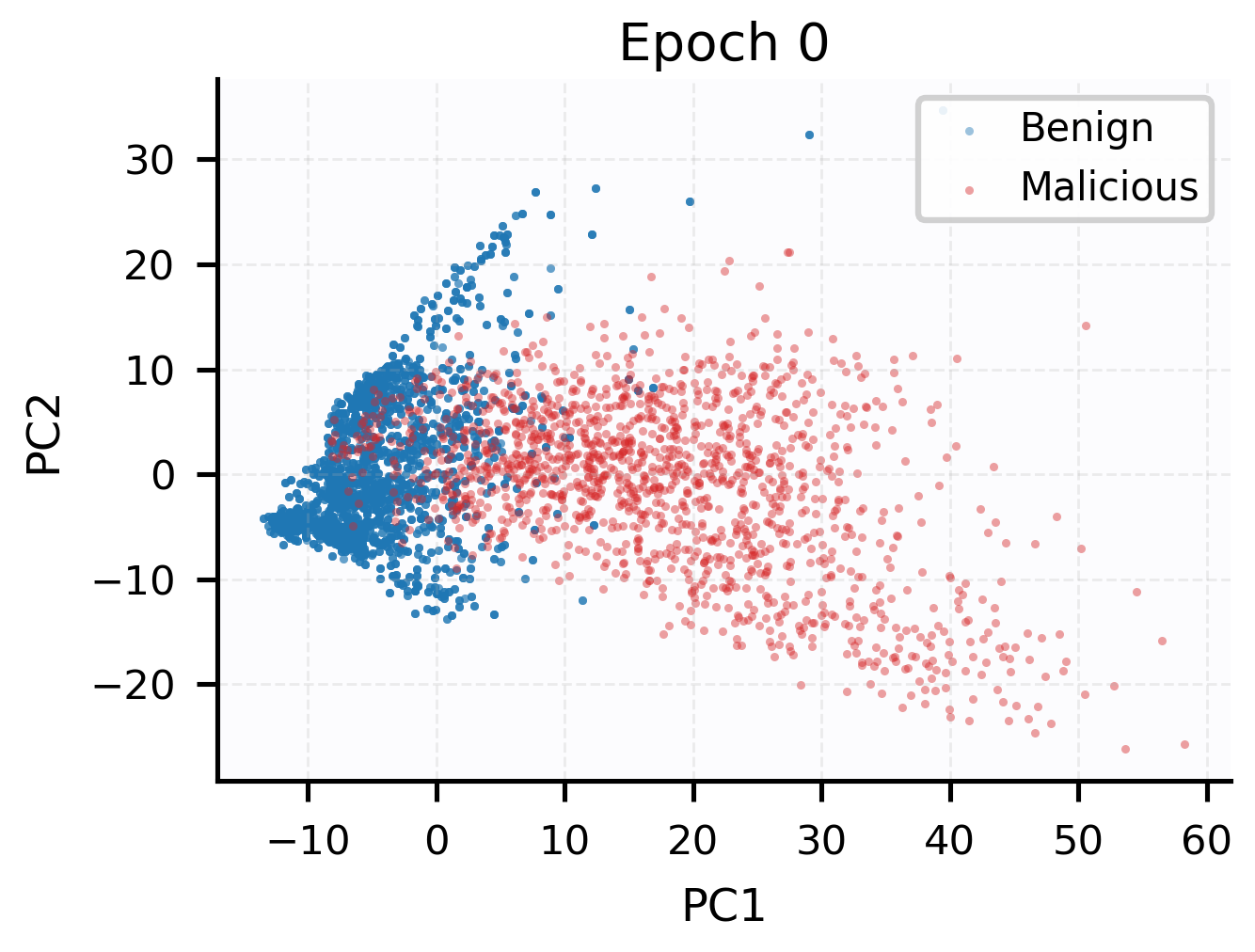}
    \hspace{0.04\linewidth}
    \includegraphics[width=0.36\linewidth]{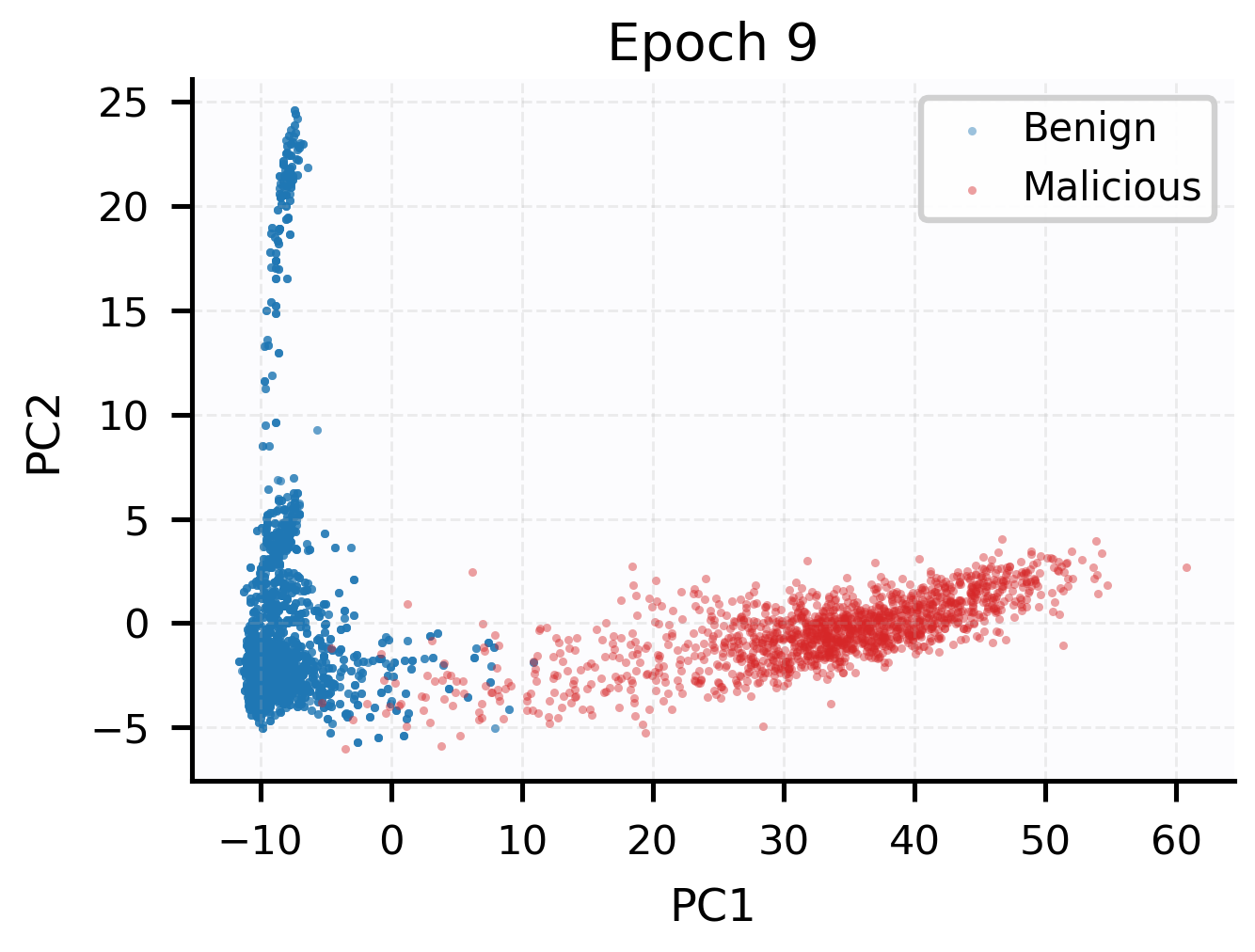}
    \caption{\textbf{PCA scatter of residual embeddings at epoch~0 (left) and epoch~9 (right),} evaluated on CP-GuardBench with $\Delta=0.25$ pooling all five attack types. Benign and malicious samples are heavily mixed at epoch~0, but form clearly separated, compact clusters by epoch~9, demonstrating convergence of the learned feature centers.}
    \label{fig:center_scatter_epochs}
    \vspace{-5mm}
\end{figure*}

To examine the stability of the learned feature centers, we visualize PCA-projected residual embeddings (pooled from all five attacks) at epoch~0, and epoch~9. As shown in Figure~\ref{fig:center_scatter_epochs}, at epoch~0 benign and malicious samples are substantially mixed with no clear separation. By epoch~9, the benign cluster (blue) becomes compact and concentrated, while malicious samples (red) are pushed into a clearly separated region. This confirms that DCCLoss drives stable and well-separated feature centers as training progresses.

\subsubsection{Visualization} Figure \ref{fig:qualitative_results} visualizes the CP system's results with and without the CP-Guard+ defense mechanism. The top row shows that, without defense, the system produces many unmatched predicted bounding boxes, i.e., false positives under the standard IoU-based detection criterion used in AP evaluation. The bottom row shows that CP-Guard+ suppresses these attack-induced false alarms while preserving correct detections.

\section{Limitation and Future Work}
\label{sec:limitation}
The CP-GuardBench dataset focuses on specific attack scenarios and may not cover all potential adversarial strategies, limiting the robustness of CP-Guard+ against entirely new types of attacks. This work focuses on representative collaborative perception attacks under a fixed attack pipeline, and does not consider fully adaptive or multi-step adversaries explicitly optimized to evade the proposed detector. In addition, the method relies on residual latent features to distinguish malicious agents, but highly dynamic and complex environments may introduce noise and false positives that affect detection accuracy. In the future, we plan to expand CP-GuardBench to cover more attack scenarios, model backbones, and parameter settings to enhance state-of-the-art evaluation, gradually making it more comprehensive and fostering the development of secure collaborative perception. Moreover, we will explore more effective feature-level defense methods to further improve the performance of malicious agent detection.

\section{Conclusion}
\label{sec:conclusion}

In this paper, we have proposed a new paradigm for malicious agent detection in CP systems, which directly detects malicious agents at the feature level without generating multiple hypothetical results, significantly reducing system complexity and computation cost.
We have also constructed a new benchmark dataset, CP-GuardBench, for malicious agent detection in CP systems, which is the first benchmark dataset in this field.
Furthermore, we have developed CP-Guard+, a resilient framework for detecting malicious agents in CP systems, capable of identifying malicious agents at the feature level without the need to verify the final perception results, significantly reducing the computation cost and improving processing throughput. We hope the new dataset and the proposed framework CP-Guard+ can contribute to the community and promote further development of secure and reliable CP systems.

{
  \footnotesize
  \bibliographystyle{unsrtnat}
  \bibliography{ref, ref2}
}

\end{document}